\documentclass[twocolumn,aps,prl,superscriptaddress]{revtex4-2}
\usepackage{color}
\usepackage{mathrsfs}
\usepackage{bm}
\usepackage{amsmath}
\usepackage{amssymb}
\usepackage{graphicx}
\usepackage{hyperref}
\usepackage{color}
\usepackage{siunitx}
\allowdisplaybreaks[3]

\makeatletter
\hypersetup{hypertex=true, colorlinks=true, linkcolor=blue, anchorcolor=blue,citecolor=blue}
\usepackage{dcolumn}
\usepackage{bm}
\usepackage{mathrsfs}
\usepackage{amsfonts}
\usepackage{xcolor}
\usepackage{epstopdf}

\begin{document}
\title{Finite-Momentum Pairing State in Unconventional Rashba Systems } 
\author{Ran Wang}
\affiliation{Anhui Provincial Key Laboratory of Low-Energy Quantum Materials and
		Devices, High Magnetic Field Laboratory, HFIPS, Chinese Academy
		of Sciences, Hefei, Anhui 230031, China}
\affiliation{Science Island Branch of Graduate School, University of Science and
		Technology of China, Hefei, Anhui 230026, China}
\author{Song-Bo Zhang}
\affiliation{Hefei National Laboratory, Hefei, Anhui, 230088, China}
\affiliation{International Center for Quantum Design of Functional Materials (ICQD),
University of Science and Technology of China, Hefei, Anhui 230026, China}
\author{Ning Hao}
\email{haon@hmfl.ac.cn}
\affiliation{Anhui Provincial Key Laboratory of Low-Energy Quantum Materials and
		Devices, High Magnetic Field Laboratory, HFIPS, Chinese Academy
		of Sciences, Hefei, Anhui 230031, China}
\affiliation{Science Island Branch of Graduate School, University of Science and
		Technology of China, Hefei, Anhui 230026, China}
	
	\begin{abstract}
		 In systems with unconventional Rashba bands, we propose that a finite-momentum pairing state can emerge without the need for an external magnetic field. We analyze the phase transition from a zero-momentum Bardeen-Cooper-Schrieffer (BCS) state to a finite-momentum pairing state using a microscopic interaction model.  We demonstrate that the coexistence of zero-momentum and Larkin-Ovchinnikov (LO)-type pairings in different channels provides a well understanding of recent experimental observations of pair-density-wave (PDW) states. Furthermore, we propose that an s-wave BCS superconductor-unconventional Rashba metal (SC-URM) junction can generate an LO-type pairing state via an orbital-selective proximity effect. This nontrivial state can be detected by measuring the Josephson current in an SC-URM-SC junction or through Josephson scanning tunneling microscopy/spectrosscopy (JSTM/S). Our results reveal that the internal multi-orbital degrees of freedom play a crucial role in facilitating finite-momentum pairing states.
	\end{abstract}
	\maketitle
	
\begin{figure}[htbp]
		\begin{center}
			\includegraphics[width=1.0\columnwidth]{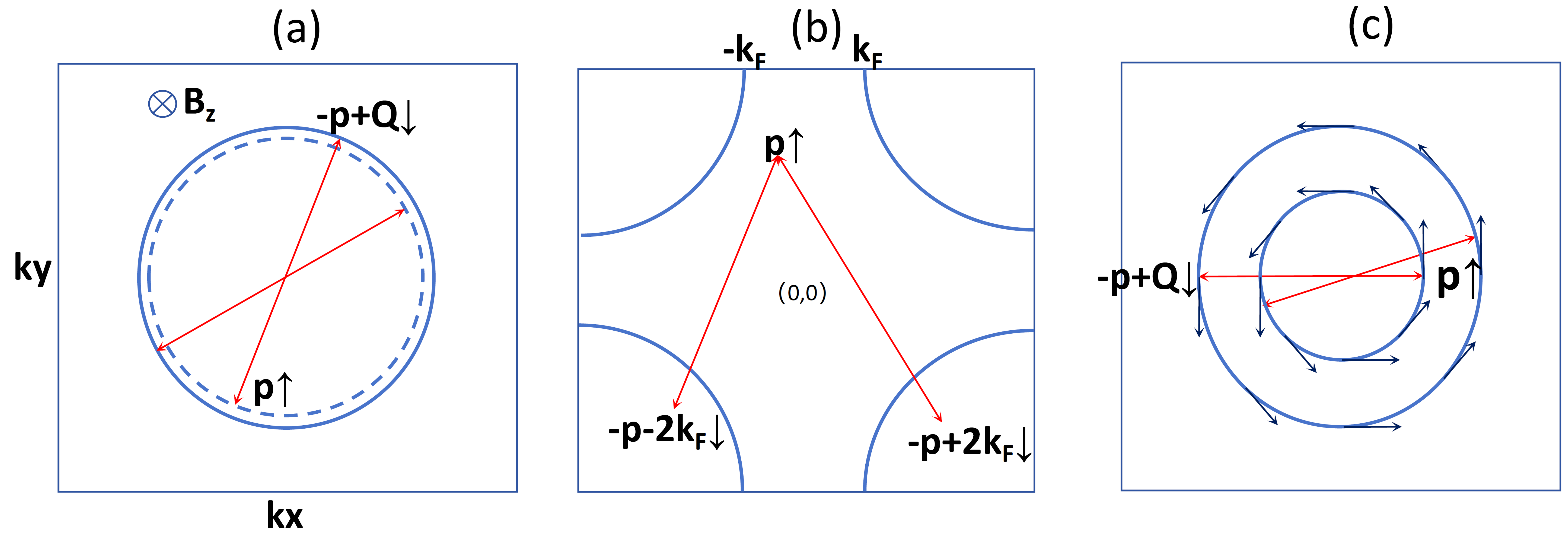}
		\end{center}
		\caption{Schematic of three types of finite-momentum pairing states. (a) The FFLO state driven by an external magnetic field. (b) The PDW state with Amperean pairing in cuprates \cite{lee2014amperean}. (c) The finte-momentum pairing state in unconventional Rashba systems with anomalous spin textures. \label{fig-fmp}}
	\end{figure}

	\textit{Introduction.}--- Finite-momentum pairing states \cite{InhomogeneoussuperconductivityincondensedmatterandQCD, annurev:/content/journals/10.1146/annurev-conmatphys-031119-050711} deviate from the conventional BCS state, in which Cooper pairs have zero total momentum. In these unconventional states, Cooper pairs acquire a finite center-of-mass momentum, resulting in spatially modulated superconducting order parameters. Two prominent examples of such states are the Fulde-Ferrell-Larkin-Ovchinnikov (FFLO) state \cite{PhysRev.135.A550, LO, kinnunen2018fulde} and the PDW state \cite{PDW, PhysRevB.79.064515, PhysRevB.81.020511, PhysRevB.89.165126, PhysRevLett.99.127003, lee2014amperean, berg2009charge,wang2018pair}. In both cases, the superconducting order parameter varies periodically in space, though with different characteristic wavelengths. This difference arises from distinct underlying mechanisms. The FFLO state, first proposed in the 1960s, occurs in systems with Zeeman splitting under an external magnetic field and has recently been suggested to arise in systems with more general time-reversal-symmetry-breaking effects, such as altermagnets \cite{altermagnets, PhysRevB.110.L060508, sim2024pair, chakraborty2024perfect} and flat-band systems \cite{sun2024flatbandfuldeferrelllarkinovchinnikovstatequantum}. These systems typically feature an imbalance between spin-up and spin-down electron populations or mismatched Fermi surfaces, leading to a tendency for the formation of an FFLO state. The finite momentum corresponds to the imbalanced or mismatched Fermi vector, as illustrated in Fig. \ref{fig-fmp}(a). However, the FFLO state is often difficult to stabilize due to its sensitivity to impurities \cite{10.1143/PTP.43.27, PhysRevB.75.184515} and the orbital pair-breaking effect of the magnetic field \cite{PhysRevLett.16.996, PhysRevB.68.184510}. In contrast, the PDW state does not require an external magnetic field, but it can coexist with or compete against other ordered states, such as charge or spin density waves, as shown in Fig. \ref{fig-fmp}(b). Systems exhibiting PDW states often have strong electronic correlations, such as high-temperature superconductors \cite{PDW, PhysRevB.79.064515, PhysRevB.81.020511, PhysRevB.89.165126, PhysRevLett.99.127003, lee2014amperean, berg2009charge,wang2018pair} and flat-band systems \cite{chen2023pair, jiang2023pair, wang2024quantum}. The complexity of these interactions makes the underlying mechanism of the PDW state a subject of ongoing debate \cite{annurev:/content/journals/10.1146/annurev-conmatphys-031119-050711, PhysRevB.79.064515, PhysRevLett.93.187002}.

Rashba spin-orbit coupling (RSOC) plays a fascinating role in the study of finite-momentum pairing states \cite{kinnunen2018fulde, PhysRevB.75.064511, PhysRevB.76.014522, Topologicalmetalsandfinite-momentumsuperconductors, Supercurrentdiodeeffectandfinite-momentumsuperconductors}. RSOC arises from structural inversion asymmetry and results in a momentum-dependent splitting of spin bands \cite{EIRashba, PismaZhETF.39.66, PhysRevB.48.8918, PhysRevB.62.4245}, which can profoundly impact the nature of superconducting pairing \cite{PhysRevLett.87.037004, PhysRevLett.92.097001}. Recently, multi-orbital degrees of freedom have been introduced and emphasized in RSOC systems \cite{PhysRevB.109.195419, song2021giant}, leading to unconventional Rashba bands with anomalous spin textures. These unconventional Rashba systems can host exotic quantum phases and phenomena, both in superconductivity \cite{wang2024superconductivitytwodimensionalsystemsunconventional} and in spintronics \cite{PhysRevB.109.195419, bhattacharya2024electricfieldinducedsecondorder, gao2018possible, song2021giant}.  

In this work, we explore the theoretical framework and propose an experimental scheme for finite-momentum pairing in unconventional Rashba systems, focusing on the conditions necessary for the emergence of such states and the potential experimental signatures that could confirm their existence. Unlike conventional FFLO states, finite-momentum pairing can arise in unconventional Rashba systems without the need for an external field. We begin by analyzing the phase transition from a BCS state to a finite-momentum pairing state using a microscopic interaction model. Our findings highlight the critical role of internal multi-orbital degrees of freedom in generating intrinsic finite-momentum states. Additionally, we show that the coexistence of zero-momentum and LO-type pairings in different channels provides a comprehensive explanation of recent experimental observations of PDW states. Furthermore, we demonstrate that LO-type pairing can be induced in an URM via an orbital-selective proximity effect and propose detecting this state through measurements of Josephson current in an SC-URM-SC junction or by employing JSTM/S.
    
\begin{figure}[]
		\begin{center}
			\includegraphics[width=1.0\columnwidth]{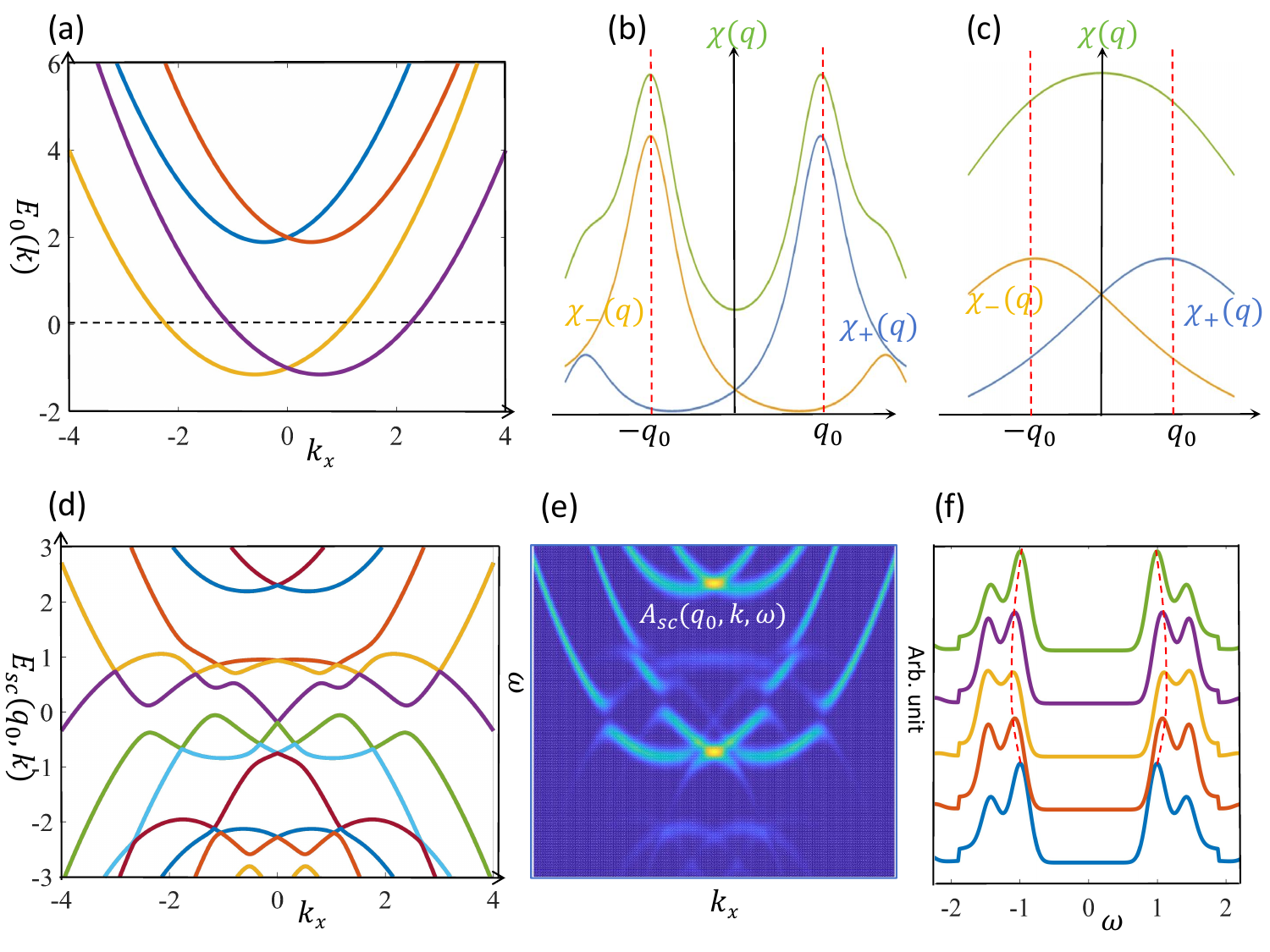}
		\end{center}
		\caption{(a) Band structures of the unconventional Rashba system. The relevant Fermi circles are shown in Fig. \ref{fig-fmp}(c). (b) and (c) Two types of superconducting susceptibility $\chi_{1u}$ in $1u$ channel under different $\lambda_{R}$. $\lambda_{R}$=0.5 in (b) and $\lambda_{R}$=0.1 in (c). Other parameters in (b) and (c) are $\gamma_0$=0.5, $\mu=1$, $\lambda$=1.5, $\lambda_{R}$=0.5, $\varepsilon $=1, and $1/k_{B}T_{c}$=5. (d) Energy dispersion of quasi-particle for pure $1u$ LO-type state. The strength of order parameter is $\Delta_{1u}$=0.5. (e) The spectral function corresponding to (d). (f) The density of states of quasi-particle spectrum for the coexistence of both zero-momentum $1g1$ pairing and LO-type $1u$ pairing. The strengths of order parameters are $\Delta_{1g1}$=1 and $\Delta_{1u}$=0.5. From bottom to top, the plots correspond to the different positions of $\bm{r}$, at which, the local LO pair is modulated according to $\Delta_{1u}\cos(\widetilde{\bm{Q}} \cdot \bm{r})$. The red dashed curves indicates the evolution of the coherent peak at different positions. In (d)-(f), other parameters are the same as those in (b). \label{fig-fflo}}
	\end{figure}

\textit{Model and pairing symmetry.}--- To describe the bands of URM \cite{PhysRevB.109.195419, wang2024superconductivitytwodimensionalsystemsunconventional}, we start with a 4$\times$4 Hamiltonian defined in the basis $\hat{c}_{\boldsymbol{k}}^{\dagger}=\left(c_{\boldsymbol{k},1\uparrow}^{\dagger},c_{\boldsymbol{k},1\downarrow}^{\dagger},c_{\boldsymbol{k},2\uparrow}^{\dagger},c_{\boldsymbol{k},2\downarrow}^{\dagger}\right)$, 
	\begin{equation}
		\hat{H}^{UR}(\boldsymbol{k})=\xi_{\boldsymbol{k}}\sigma^{0}\tau^{0}-\lambda_{R}(k_{y}\sigma^{1}-k_{x}\sigma^{2})(\tau^{0}+\varepsilon\tau^{1})+\lambda\sigma^{3}\tau^{2}.\label{H0}
	\end{equation}
Here, $i\in\{1,2\}$ denotes the effective orbitals, and $\sigma\in\{\uparrow,\downarrow\}$ reperents the spin
	in the electron annihilation operator $c_{\bm{k},i\sigma}$. The Pauli matrices
	$(\sigma^{0},\bm{\sigma})$ and $(\tau^{0},\bm{\tau})$ span the spin
	and orbital spaces, respectively. $\xi_{\boldsymbol{k}}=\gamma_0k^2-\mu_{0}$, where $\gamma_0$ controls the band curvature near the Fermi surface, and $\mu_{0}$ is the chemical potential. For convenience, we rescale the chemical potential as $\mu=\mu_{0}+\lambda$. 
	$\lambda_{R}$ reperents the strength of RSOC, while $\lambda$ denotes the strength of on-site SOC. $\varepsilon$ scales the relative strength of inter-orbital RSOC. The energy dispersion of (\ref{H0}) is $E_{\alpha\beta}=\xi_{\boldsymbol{k}}+\alpha\sqrt{\lambda^{2}+\lambda_{R}^{2}k^{2}}+\beta\varepsilon\lambda_{R}k$, where $\alpha,\beta\in\{+,-\}$ labels the band indices. When the chemical potential lies within the band gap at $k=0$, the two Fermi surfaces corresponding to the lower two bands share the same spin chirality, forming the unconventional Rashba bands, as illustrated in Figs. \ref{fig-fmp}(c) and \ref{fig-fflo}(a).

The classification of possible superconducting pairings has been previously established \cite{wang2024superconductivitytwodimensionalsystemsunconventional}. Here, we focus on pairings arising from widely-used on-site attractive interactions, summarized in Table \ref{pair}. In systems with $C_{4v}$ symmetry, the $1g1$ and $1g2$ pairings belong to the $A_1$ irreducible representation (IR), while the $1u$ pairing belongs to the $A_2$ IR. These IRs indicate whether coupling occurs between different pairing channels, consistent with our explicit calculations.
	
	\textit{Intrinsic finite-momentum pairing.}---  Since we are interested in finite-momentum pairings, there should be a significant splitting between the lower two bands. To achieve this, we relax the approximation ($\lambda_{R}k_F/\lambda\ll1$) used in our previous work \cite{wang2024superconductivitytwodimensionalsystemsunconventional}. To make the model tractable, we consider the case where $\lambda_{R}k_F/\lambda<1$ but not too small. Intuitively, the mismatched concentric Fermi circles may induce finite-momentum Cooper pairs, akin to the magnetic-field-induced FFLO state, as illustrated in Fig. \ref{fig-fmp}(a) and \ref{fig-fmp}(c). 
		\begin{table}[ptb]
		\caption{Possible IRs of superconducting pairings in URM under the constraints of group $C_{4v}$ and on-site attractive interactions.}
		\label{pair}
		\begin{ruledtabular}
			\begin{tabular}{ccc}
				Label & IRs for $C_{4v}$ & Pairing form $\hat{\Gamma}$\tabularnewline
				\hline 
				$1g1$ & $A_1$ & $i\sigma^{2}\tau^{0}$\tabularnewline
				$1g2$ & $A_1$ & $i\sigma^{2}\tau^{1}$\tabularnewline
				$1u$ & $A_2$ & $i\sigma^{2}\tau^{3}$\tabularnewline
			\end{tabular}
		\end{ruledtabular}
	\end{table}

Next, we explore the possibility of finite-momentum pairing using a phenomenological theory of unconventional superconductivity \cite{AnunusualsuperconductivityinUBe13, Superconductingclassesinheavy-fermionsystems, P-WaveSuperconductivityinCubicMetals, RevModPhys.63.239, wen2022superconducting, yang2020soft}. 
The microscopic Hamiltonian is given by $H_{SC}=H_{UR}+H_{int}$, where
	\begin{align}
		H_{UR}=&\sum_{\boldsymbol{k}}\hat{c}_{\boldsymbol{k}}^{\dagger}\hat{H}^{UR}(\boldsymbol{k})\hat{c}_{\boldsymbol{k}}. \\
		H_{int}=&-\frac{1}{2}\sum_{\boldsymbol{k},\boldsymbol{k'},\Gamma,\boldsymbol{q}}v_{\Gamma}\hat{c}_{\boldsymbol{k}}^{\dagger}\hat{\Gamma}\hat{c}_{-\boldsymbol{k}+\boldsymbol{q}}^{*}\hat{c}_{-\boldsymbol{k'}+\boldsymbol{q}}^{T}\hat{\Gamma}^{\dagger}\hat{c}_{\boldsymbol{k'}}. \label{Hint}
	\end{align}
Here, $\Gamma\in\{1g1,1g2,1u\}$, and $v_{\Gamma}$ takes constant values. For simplicity, we search for states with $\boldsymbol{q_k}=q_0\hat{\boldsymbol{k}}$. From the linearized gap equations near the superconducting transition temperature $T_c$, we find that zero-momentum Cooper pairs are always favored in the $1g1$ and $1g2$ states \cite{sm}. However, in the $1u$ state, finite-momentum pairing ($q_0 = \varepsilon\lambda_{R}/\gamma_0 '$) can be favored if the band splitting is sufficiently large under the condition $q_0/k_0>(\gamma_0 'k_0^2/2/\pi k_B T_c)^{-1}/a_0$ with $a_0\approx3.3$, $\gamma_0 '=\gamma_0-\lambda_{R}^2/(2\lambda)$. Otherwise, the trivial zero-momentum pairing is preferred. Note that $q_0\equiv k_{out}-k_{in}=\varepsilon\lambda_{R}/\gamma_0 '$, $k_0\equiv k_{out}+k_{in}=D_0/\gamma_0 '$ with $D_0=\sqrt{\varepsilon^2\lambda_{R}^2+4\gamma_0 '\mu}$, $k_{in(out)}\approx(D_0-(+)\varepsilon\lambda_{R})/2\gamma_0 '$. The transition between finite- and zero-momentum pairing is determined by the superconducting susceptibility $\chi_{1u}$ as a function of $q$ under different parameters, as shown in Fig. \ref{fig-fflo} (b) and (c). The peaks of $\chi_{1u}$ indicates the presence of supperconductivity therein.     

The finite-momentum pairing here differs from the FFLO pairing, which has a well-defined center-of-mass momentum $\boldsymbol{q_0}$ for Cooper pairs \cite{PhysRevLett.96.060401, SHEEHY20071790, annurev:/content/journals/10.1146/annurev-conmatphys-031119-050711}. However, due to $C_{4v}$ symmetry, only four independent $\boldsymbol{q_0}$ directions remain. This symmetry allows for a variety of finite-momentum pairings, such as FF-type, LO-type, bidirectional, or double-helix pairings \cite{PhysRevB.89.165126}. In practice, perturbations that break the $C_4$ rotational symmetry make the LO-type pairing the most likely \cite{note1}. Figure \ref{fig-fflo} (d) and 2(e) show the quasi-particle spectrum and spectral function for a superconducting state with pure LO-type pairing, revealing broken particle-hole symmetry and a pair of residual Bogoliubov quasi-particle nodal points.   

\textit{Implication to experiments.}--- Recent experiments have reported signatures of PDW orders in monolayer Fe(Te,Se), EuRbFe$_{4}$As$_{4}$ and UTe$_{2}$ \cite{liu2023pair, zhao2023smectic, wei2024observation, kong2024observation, zhang2024visualizing, gu2023detection}. Seizing the multi-orbital and inversion-symmetry-breaking characteristics of the systems, our findings may provide a comprehensive understanding of the phenomena observed in the iron-based systems. Specifically, our calculations reveal that the superconducting susceptibility for the $1g1$ pairing is significantly stronger than that of the $1u$ pairing, with the latter suppressed by a factor of $(\lambda_{R}k_F/\lambda)^2$ \cite{sm}. This suggests that $1g1$ pairing is dominant, while the LO-type $1u$ pairing plays a secondary role when the pairing interactions in both channels are comparable. Importantly, the coexistence of both $1g1$ and LO-type $1u$ pairings can explain the experimentally observed spatial modulation in superconducting order parameter, as shown in Fig. \ref{fig-fflo}(f). The effective superconducting gap can be expressed as $\widetilde{\Delta }_{0}+\widetilde{\Delta }_{1}\cos (\widetilde{\bm{Q}}\cdot \bm{r})$, where $\widetilde{\bm{Q}}=2\bm{q_{0}}$, owning to the orthogonality of the $1g1$ and $1u$ channels. This explains the experimentally observed short-period spatial modulation or large $\widetilde{\bm{Q}}$  of the superconducting gap due to the doubled center-of-mass momentum $\bm{q_{0}}$ here. The temperature evolution of the PDW state observed in EuRbFe$_{4}$As$_{4}$ can be explained due to the weaker LO-type $1u$ pairing relative to the $1g1$ pairing. As the temperature rises, the mixed state involving zero-momentum $1g1$ and LO-type $1u$ pairing gradually transitions into a uniform superconducting state with either both zero-momentum $1g1$ and $1u$ pairings or pure zero-momentum $1g1$ pairing. The impact of temperature effect on $1u$ pairing is shown in Ref. \cite{sm}. Furthermore, under an external magnetic field, the PDW order is suppressed within the vortex halo due to pair-breaking effects on both zero-momentum $1g1$ and LO-type $1u$ pairings from the magnetic field. However, the PDW order persists outside the vortex halo, where the effective magnetic field vanishes.

\begin{figure}[htbp]
		\begin{center}
			\includegraphics[width=1.0\columnwidth]{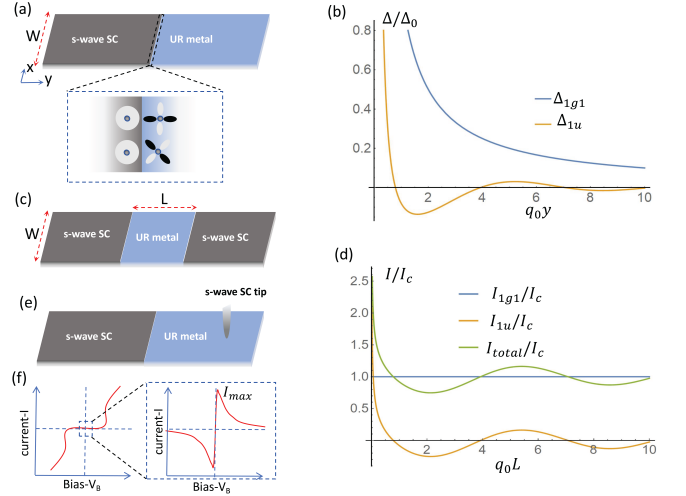}
		\end{center}
		\caption{(a) Schematic of a planar SC-URM junction. The zoom-in schematically illustrates the orbital selective proximity coupling between SC and URM on the interface. Here, we assume main s orbitals in SC and main $d_{x^2-y^2}$, $d_{xy}$ orbitals in URM. (b) The two local orbital-selective-proximity-effect-induced order parameters in the URM as a function of $q_0y$. (c) Schematic of a planar SC-URM-SC Josephson junction. The width in $x$-direction is $W$ and the length of middle URM in $y$-direction is $L$. (d) The two individual and total critical currents as a function of $q_0L$ in the planar SC-URM-SC Josephson junction. (e) Schematic of a JSTM/S platform to measure the tunneling current accross the SC tip. (f) Schematic of typical $I-V_B$ relation from JSTM/S measurement shown in (e). The zoom-in schematically illustrates the $I-V_B$ relation near zero bias $V_B$. \label{sus-stm}}
	\end{figure}

\textit{Extrinsic LO-pairing from proximity effect.}--- The coexistence of $1g1$ and finite-momentum $1u$ pairings imposes restrictions on the pairing interactions, especially when these interactions differ in intra-orbital channels. This suggests that multi-orbital degrees of freedom are crucial for realizing finite-momentum pairing. However, when the orbital nature is disregarded, the pairings in all the three IRs reduce to s-wave spin-singlet pairings. This implies that these pairings can arise through the proximity effect between an URM and a conventional SC. To illustrate this, we consider a planar URM-SC junction, as depicted in Fig. \ref{sus-stm}(a). At the interface, the proximity effect can be described by a mean-field Hamiltonian,
	\begin{align}
		H_{prox}=&H_{UR}+\sum_{\boldsymbol{k},\sigma}\xi_{\boldsymbol{k}}^dd_{\boldsymbol{k}\sigma}^{\dagger}d_{\boldsymbol{k}\sigma}-\sum_{\boldsymbol{k}}\left(\Delta d_{\boldsymbol{k}\uparrow}^{\dagger}d_{-\boldsymbol{k}\downarrow}^{\dagger}+h.c.\right) \notag \\ &+\frac{1}{2}\sum_{\boldsymbol{k},\boldsymbol{q},i,\sigma}\left(\mathbf{t}_ic_{\boldsymbol{k+\frac{q}{2}},i\sigma}^{\dagger}d_{\boldsymbol{k}\sigma}+h.c.\right). \label{prox}
	\end{align}
Here, $d_{\boldsymbol{k}\sigma}^{\dagger}$ ($d_{\boldsymbol{k}\sigma}$) are the electron creation (annihilation) operators in the SC. $\xi_{\boldsymbol{k}}^d$ denotes the energy dispersion for the normal states of the SC, while $\Delta$ refers to the pairing order parameter. The second line describes the couplings between the URM and SC. For simplicity, we assume that the hopping integral  $\mathbf{t}_{\boldsymbol{k}\boldsymbol{p}}\equiv \mathbf{t}_i$ are independent of $\boldsymbol{k}$ and $\boldsymbol{p}\equiv\boldsymbol{k}+\boldsymbol{q}/2$. Additionally, we consider orbital-selective hopping integrals ($\mathbf{t}_1$, $\mathbf{t}_2$). Here, $\boldsymbol{q}$ represents the center-of-mass momentum of the Cooper pairs, which is determined by the Cooper-pair propagator in each channel as they propagate into the URM. The coupling between the URM and SC mixes electrons and holes, revealing the nature of the induced order parameter in the URM. This interfacial pairing, induced by the proximity effect, can be derived using either the down-folding perturbation method or the Green-function equations of motion method \cite{sm}. Assuming a weak proximity effect ($|\mathbf{t}_{1/2}|/\Delta\ll1$), we obtain three effective orbital-selective order parameters ($\Delta_{eff}^{1g1}(\boldsymbol{k}), \Delta_{eff}^{1g2}(\boldsymbol{k}), \Delta_{eff}^{1u}(\boldsymbol{k}))\propto - ((\mathbf{t}_{1}^{2}+\mathbf{t}_{2}^{2}),2\mathbf{t}_{1}\mathbf{t}_{2},(\mathbf{t}_{1}^{2}-\mathbf{t}_{2}^{2}))/\Delta $.
    
Different interfacial pairings can be realized by tuning the hopping intergrals. Specifically, we consider the case involves two orbitals of the URM, primarily derived from $d_{x^2-y^2}$ and $d_{xy}$ orbitals. Orbital-selective proximity effects are realized by aligning the interface along either $x$ or $y$ axes, or along the diagonals $y=\pm x$, as shown in Fig. \ref{sus-stm}(a). This configuration results in $\mathbf{t}_1\neq0,\mathbf{t}_2=0$ or vice versa, causing the $1g2$ channel to vanish while the $1g1$ and $1u$ channels coexist. 
    
The orbital-selective interfacial order parameters penetrate into the URM with a decay behavior. To describe the propagation of the order parameter, we introduce the Cooper-pair propagator \cite{altermagnets, sm, PhysRevB.5.923},
    \begin{equation}
    	\mathcal{D}_{\Gamma,\Gamma'}(1;2)=-\langle T_{\tau}\hat{c}^{T}(1)\hat{\Gamma}^{\dagger}\hat{c}(1)\hat{c}^{\dagger}(2)\hat{\Gamma}'\hat{c}^*(2)\rangle. \label{propagator}
    \end{equation}
    Here, index $i\in\{1,2\}$ labels real space coordinate $\boldsymbol{r_i}=(x_i,y_i)$ and imaginary time $\tau_i$, that $\hat{c}^{\dagger}(i)\equiv\hat{c}^{\dagger}(\boldsymbol{r_i},\tau_i)=\left(c_{1\uparrow}^{\dagger}(\boldsymbol{r_i},\tau_i),c_{1\downarrow}^{\dagger}(\boldsymbol{r_i},\tau_i),c_{2\uparrow}^{\dagger}(\boldsymbol{r_i},\tau_i),c_{2\downarrow}^{\dagger}(\boldsymbol{r_i},\tau_i)\right)$. For convenience, we define $\boldsymbol{r}=\boldsymbol{r_1}-\boldsymbol{r_2}$ and  $\boldsymbol{\tau}=\boldsymbol{\tau_1}-\boldsymbol{\tau_2}$. $\Gamma$ and $\Gamma'$ stand for pairings in $\{1g1,1g2,1u\}$ in Table \ref{pair}. $T_{\tau}$ is the time ordering operator. After performing a Fourier transformation in imaginary time, the Cooper-pair propagator can be calculated \cite{sm}.
    
    The local order parameter inside the URM is phenomenologically derived from the propagators as,
    \begin{equation}
    	\Delta_{\Gamma}(\boldsymbol{r_1})=\sum_{\Gamma'}\lambda_{\Gamma'}\int_{-W/2}^{W/2} dx_2D_{\Gamma,\Gamma'}(x_1,y_1;x_2,0). \label{op}
    \end{equation}
    Here, the Cooper-pair propagator takes the retarded form $D_{\Gamma,\Gamma'}(\boldsymbol{r_1};\boldsymbol{r_2})\equiv D_{\Gamma,\Gamma'}(\boldsymbol{r_1}-\boldsymbol{r_2},\omega\to0^+)$. $\lambda_{\Gamma'}$ is the strength of the order parameter for pairing form $\Gamma'$ at the interface. Since the proximity effect is orbital selective with $\mathbf{t}_1\neq0$ and $\mathbf{t}_2=0$ as shown in Fig. \ref{sus-stm}(a), $\lambda_{\Gamma'}$ vanishes for the $1g2$ pairing, leaving only the $1g1$ and $1u$ pairings. The propagators the for $1g1$ and $1u$ channels are given by, 
\begin{eqnarray}
D_{1g1}(\boldsymbol{r}) &=&-1/2\pi ^{2}\gamma_0^{\prime }r^{2},  \label{cpp-1g1}
\\
D_{1u}(\boldsymbol{r}) &=&-\frac{g(+,-)}{4\pi ^{2}\gamma_0^{\prime }r^{2}}\eta\cos\left(q_0r\right).
\label{cpp-1u}
\end{eqnarray}%
Here, $g(+,-)=1-\cos\phi_+\cos\phi_-+\sin\phi_+\sin\phi_-$, $\phi_{\pm}=\arctan\left[\lambda_{R}(D_0\mp\varepsilon\lambda_{R})/2\gamma_0 '\lambda\right]$, $\eta = \sqrt{\left(1-\varepsilon ^{2}\frac{\lambda _{R}^{2}}{D_{0}^{2}}\right)}$. 
Applying this to the SC-URM junction in Fig. \ref{sus-stm}(a), both $1g1$ and LO-type $1u$ pairings can be induced by the proximity effect. For a wide planar URM-SC junction, the induced order parameters become homogeneous along the interface and independent on the junction width $W$, i.e., $\Delta( \boldsymbol{r_1})\approx \Delta(y_1)=\Delta(y)$. Combining with Eqs. (\ref{op}), (\ref{cpp-1g1}) and (\ref{cpp-1u}), we have 
\begin{align}
	\Delta_{1g1}(y)&=\varepsilon\Delta_0/{q_0y}, \\
	\Delta_{1u}(y)&=\varepsilon\Delta_0\frac{g(+,-)}{\sqrt{2\pi }(q_0y)^\frac{3}{2}}\eta\cos(q_0y+\frac{\pi}{4}).
\end{align}
Here, $\Delta_0=\lambda_0\lambda_{R}/2\pi \gamma_0 '^2$ under the assumption of $\lambda_{1g1}=\lambda_{1u}=\lambda_0$. $\Delta_{1g1}(y)$ and $\Delta_{1u}(y)$ exhibit spatial decay proportional to $y^{-1}$ and $y^{-3/2}$, respectively. Interestingly, $\Delta_{1u}(y)$ shows additional cosine oscillatory behavior, indicating a nonzero center-of-mass momentum $q_0$, corresponding to the difference in the Fermi wave vectors between the inner and outer Fermi surfaces.
Notably, $g(+,-)\sim \frac{1}{2}(\lambda_{R}D_0/\gamma_0\lambda)^2 \approx (\lambda_{R}k_F/\lambda)^2$, meaning the $1u$ pairing is suppressed when $\lambda_{R}k_F\ll\lambda$. This is consistent with our earlier microscopic results. In Fig. \ref{sus-stm}, we plot $\Delta(0,y)$ as a function of dimensionless length $q_0y$, clearly showing both the oscillation and decay behavior for $1u$ pairing.
 
    \textit{Detection of LO-type pairing.}--- To detect the proximity-effect-induced LO-type pairing, one can exploit the Josephson effect \cite{altermagnets, ControlledfinitemomentumpairingandspatiallyvaryingorderparameterinproximitizedHgTequantumwells, FinitemomentumCooperpairinginthree-dimensionaltopologicalinsulatorJosephsonjunctions}. To derive the current, we begin with the action for the URM under the proximity effect of the SC. In the real space and frequency domains, the action is given by,
\begin{align}
S& =-\sum_{i\omega _{n}}\int d^{2}r_{1}d^{2}r_{2}\hat{\bar{c}}(\boldsymbol{%
r_{1}};i\omega _{n})\mathcal{G}^{-1}(\boldsymbol{r};i\omega _{n})\hat{c}(%
\boldsymbol{r_{2}};i\omega _{n})  \nonumber \\
-& \sum_{i\omega _{n},\Gamma }\int d^{2}r_{1}\left[ \Delta _{\Gamma }(%
\boldsymbol{r_{1}})\hat{\bar{c}}(\boldsymbol{r_{1}};i\omega _{n})\hat{\Gamma}%
\hat{\bar{c}}^{T}(\boldsymbol{r_{1}};-i\omega _{n})+h.c.\right] .
\end{align}%
    Here, the spinor of Grassman numbers are $\hat{\bar{c}}=\left(\bar{c}_{1\uparrow},\bar{c}_{1\downarrow},\bar{c}_{2\uparrow},\bar{c}_{2\downarrow}\right)$. $\mathcal{G}(\boldsymbol{r};i\omega_n)$ is the relevant Matsubara Green function with the frequency $i\omega_n$ of fermions and integer $n$. \cite{wang2024superconductivitytwodimensionalsystemsunconventional}. $\Delta_{\Gamma}$ is the order parameter for  $\hat{\Gamma}\in\{1g1,1g2,1u\}$. Integration of the electron field gives a free energy in the second order of $\Delta_\Gamma$. 
    \begin{equation}
    	F_2=-2\sum_{\Gamma,\Gamma'}\int d^2r_1d^2r_2\Delta_{\Gamma}(\boldsymbol{r_2})\Delta^*_{\Gamma'}(\boldsymbol{r_1})D_{\Gamma,\Gamma'}(\boldsymbol{r_1};\boldsymbol{r_2}). \label{gp}
    \end{equation}

We consider a planar SC-URM-SC junction in the $x$-$y$ plane to compute the Josephson current. The junction has length $L$ in $y$-direction and width $W$ in $x$-direction, as shown in Fig. \ref{sus-stm}(c). Considering a weak proximity effect, $\Delta_{\Gamma}(\boldsymbol{r})$ is nonzero only at the URM-SC interface, i.e., $\Delta_{\Gamma}=\Delta_{\Gamma,a}\delta(y)+\Delta_{\Gamma,b}\delta(y-L)$ with $\delta(y)$ a Delta function. Since the interfacial order parameters in different channels are induced by the same SC, they share the same phase $\Delta_{\Gamma,j}=\lambda_{\Gamma,j}e^{i\phi_j}$ with $j\in\{a,b\}$, and $\lambda_{\Gamma,j}$ is real. Due to the orbital-selective proximity effect, the cross terms $D_{\Gamma,\Gamma'}$ ($\Gamma\neq\Gamma'$) in Eq. (\ref{gp}) do not contribute. The Josephson current is given by $I=e/\hbar\partial F_2(\delta\phi)/\partial\delta\phi$, leading to $I=\sin\left(\delta\phi\right)\sum_{\Gamma}I_{\Gamma}$, where the critical current for each channel is given by 
    \begin{equation}
    	I_{\Gamma}=4\frac{e}{\hbar}\lambda_{\Gamma,a}\lambda_{\Gamma,b}\int_{-W/2}^{W/2} dx_1dx_2D_{\Gamma}(x_1,L;x_2,0), \label{current}
    \end{equation}
and $\delta\phi=\phi_a-\phi_b$. This result indicates that the Josephson critical current is determined by the propagation of Cooper pairs. Using Eqs. (\ref{cpp-1g1}) and (\ref{cpp-1u}), we obtain
\begin{eqnarray}
I_{1g1} &\equiv &I_{c}=2(w/l)e\lambda _{a}\lambda _{b}/\pi \gamma_0^{\prime }\hbar,
\label{cu-1g1} \\
I_{1u} &=&\frac{g(+,-)}{\sqrt{2\pi l}}I_{c}\eta\cos\left( l+\frac{\pi }{4}%
\right) .  \label{cu-1u}
\end{eqnarray}
 Here, we define the dimensionless quanties $l=q_0L$, $w=q_0W$. The critical current of $1u$ channel in Eq. (\ref{cu-1u}) inherits spatial oscillations from the propagator, indicating that positive, negative or zero currents can occur by varying the length L of the URM or by introducing dopings to change $q_0$  in the junction as the decay behavior is renormalized.
The renormalized critical currents for individual and total channels are shown in Fig. \ref{sus-stm}(d).
 
    In experiments, JSTM/S provides an alternative method to detect LO-type pairing \cite{hamidian2016detection, cho2019strongly}. In the setup shown in Fig. \ref{sus-stm}(e), the local critical current can be computed using Eq. (\ref{current}) with replacing $\int dx_1dx_2\to\int dx_1dx_2\delta(x_1)$ and $L\to y$ under the conditions of large width $W$ and negligible contributions from reflected Cooper pairs. The schematic of local $I-V_B$ relation is shown in Fig. \ref{sus-stm}(f). The maximum current $I_{max}(y)$ across the SC tip at a finite bias is proportional to $[I_{1g1}(y)+I_{1u}(y)]^{2}$
 \cite{hamidian2016detection, cho2019strongly, anchenko1969josephson, ingold1994cooper}. Thus, the current $I_{max}(y)$ should exhibit a strip pattern with a spatial period of $2\pi/q_0$, as renormalized by the coexistence of the trivial $1g1$ channel, similar to Fig. \ref{sus-stm}(d).  Note that the JSTM/S method can also used to detect the LO states in pristine materials due to the similar oscillatory behavior in real space.
    
 \textit{Discussion and conclusion.}---  It is worth noting that in the usual FFLO state, disorder tends to suppress finite-momentum pairing\cite{takada1970superconductivity}. However, for our two-band model, we find that the intrinsic finite-momentum pairing has a selective response to disorder. Specifically, it is stable against intra-band scattering caused by weak disorder but is suppressed by inter-band scattering\cite{sm}. More importantly, for the LO state induced by the proximity effect, finite-momentum pairing is stable against weak disorder. This is because the LO pairing induced in the unconventional Rashba metal originates from the penetration effect of zero-momentum Cooper pairs in the superconducting regime, and its strength is entirely determined by the strength of the zero-momentum Cooper pairs in the superconducting region and the interface coupling strength.

 In summary, we demonstrate that both intrinsic and extrinsic finite-momentum pairings can be generated in unconventional Rashba systems, driven by microscopic interactions and the superconducting proximity effect, respectively. In both cases, the internal multi-orbital structure plays a crucial role. An unexpected and significant result is that the coexistence of both zero-momentum and LO-type pairings in different channels offers valuable insights into experimentally observed PDW states. We also propose two methods for detecting the LO-type pairing induced by the proximity effect. This investigation provides a deeper understanding of the rich and complex behavior of finite-momentum pairings in superconductors, particularly in multi-orbital systems.
\begin{acknowledgments}
 This work was financially
supported by the National Key R\&D Program of China (Grants No.
2022YFA1403200, 2024YFA1613200), National Natural Science Foundation of
China (Grants No. 92265104, No. 12022413), the Basic Research Program of the Chinese
Academy of Sciences Based on Major Scientific Infrastructures (Grant No. JZHKYPT-2021-08), the CASHIPS Director’s Fund (Grant No. BJPY2023A09), the \textquotedblleft
Strategic Priority Research Program (B)\textquotedblright\ of the Chinese
Academy of Sciences (Grant No. XDB33030100), Anhui Provincial Major S\&T Project(s202305a12020005) and the Major Basic Program of Natural Science Foundation of Shandong Province (Grant No. ZR2021ZD01), and
the High Magnetic Field Laboratory of Anhui Province under Contract No. AHHM-FX-2020-02. S.B.Z acknowledges the start-up fund at HFNL, the Innovation Program for Quantum Science and Technology (Grant No. 2021ZD0302800), and Anhui Initiative in Quantum Information Technologies (Grant No. AHY170000).
\end{acknowledgments}

%\bibliography{ref}

\begin{thebibliography}{65}%
\makeatletter
\providecommand \@ifxundefined [1]{%
 \@ifx{#1\undefined}
}%
\providecommand \@ifnum [1]{%
 \ifnum #1\expandafter \@firstoftwo
 \else \expandafter \@secondoftwo
 \fi
}%
\providecommand \@ifx [1]{%
 \ifx #1\expandafter \@firstoftwo
 \else \expandafter \@secondoftwo
 \fi
}%
\providecommand \natexlab [1]{#1}%
\providecommand \enquote  [1]{``#1''}%
\providecommand \bibnamefont  [1]{#1}%
\providecommand \bibfnamefont [1]{#1}%
\providecommand \citenamefont [1]{#1}%
\providecommand \href@noop [0]{\@secondoftwo}%
\providecommand \href [0]{\begingroup \@sanitize@url \@href}%
\providecommand \@href[1]{\@@startlink{#1}\@@href}%
\providecommand \@@href[1]{\endgroup#1\@@endlink}%
\providecommand \@sanitize@url [0]{\catcode `\\12\catcode `\$12\catcode
  `\&12\catcode `\#12\catcode `\^12\catcode `\_12\catcode `\%12\relax}%
\providecommand \@@startlink[1]{}%
\providecommand \@@endlink[0]{}%
\providecommand \url  [0]{\begingroup\@sanitize@url \@url }%
\providecommand \@url [1]{\endgroup\@href {#1}{\urlprefix }}%
\providecommand \urlprefix  [0]{URL }%
\providecommand \Eprint [0]{\href }%
\providecommand \doibase [0]{https://doi.org/}%
\providecommand \selectlanguage [0]{\@gobble}%
\providecommand \bibinfo  [0]{\@secondoftwo}%
\providecommand \bibfield  [0]{\@secondoftwo}%
\providecommand \translation [1]{[#1]}%
\providecommand \BibitemOpen [0]{}%
\providecommand \bibitemStop [0]{}%
\providecommand \bibitemNoStop [0]{.\EOS\space}%
\providecommand \EOS [0]{\spacefactor3000\relax}%
\providecommand \BibitemShut  [1]{\csname bibitem#1\endcsname}%
\let\auto@bib@innerbib\@empty
%</preamble>
\bibitem [{\citenamefont {Casalbuoni}\ and\ \citenamefont
  {Nardulli}(2004)}]{InhomogeneoussuperconductivityincondensedmatterandQCD}%
  \BibitemOpen
  \bibfield  {author} {\bibinfo {author} {\bibfnamefont {R.}~\bibnamefont
  {Casalbuoni}}\ and\ \bibinfo {author} {\bibfnamefont {G.}~\bibnamefont
  {Nardulli}},\ }\bibfield  {title} {\bibinfo {title} {Inhomogeneous
  superconductivity in condensed matter and qcd},\ }\href
  {https://doi.org/10.1103/RevModPhys.76.263} {\bibfield  {journal} {\bibinfo
  {journal} {Reviews of Modern Physics}\ }\textbf {\bibinfo {volume} {76}},\
  \bibinfo {pages} {263} (\bibinfo {year} {2004})}\BibitemShut {NoStop}%
\bibitem [{\citenamefont {Agterberg}\ \emph {et~al.}(2020)\citenamefont
  {Agterberg}, \citenamefont {Davis}, \citenamefont {Edkins}, \citenamefont
  {Fradkin}, \citenamefont {Van~Harlingen}, \citenamefont {Kivelson},
  \citenamefont {Lee}, \citenamefont {Radzihovsky}, \citenamefont {Tranquada},\
  and\ \citenamefont
  {Wang}}]{annurev:/content/journals/10.1146/annurev-conmatphys-031119-050711}%
  \BibitemOpen
  \bibfield  {author} {\bibinfo {author} {\bibfnamefont {D.~F.}\ \bibnamefont
  {Agterberg}}, \bibinfo {author} {\bibfnamefont {J.~S.}\ \bibnamefont
  {Davis}}, \bibinfo {author} {\bibfnamefont {S.~D.}\ \bibnamefont {Edkins}},
  \bibinfo {author} {\bibfnamefont {E.}~\bibnamefont {Fradkin}}, \bibinfo
  {author} {\bibfnamefont {D.~J.}\ \bibnamefont {Van~Harlingen}}, \bibinfo
  {author} {\bibfnamefont {S.~A.}\ \bibnamefont {Kivelson}}, \bibinfo {author}
  {\bibfnamefont {P.~A.}\ \bibnamefont {Lee}}, \bibinfo {author} {\bibfnamefont
  {L.}~\bibnamefont {Radzihovsky}}, \bibinfo {author} {\bibfnamefont {J.~M.}\
  \bibnamefont {Tranquada}},\ and\ \bibinfo {author} {\bibfnamefont
  {Y.}~\bibnamefont {Wang}},\ }\bibfield  {title} {\bibinfo {title} {The
  physics of pair-density waves: Cuprate superconductors and beyond},\ }\href
  {https://doi.org/https://doi.org/10.1146/annurev-conmatphys-031119-050711}
  {\bibfield  {journal} {\bibinfo  {journal} {Annual Review of Condensed Matter
  Physics}\ }\textbf {\bibinfo {volume} {11}},\ \bibinfo {pages} {231}
  (\bibinfo {year} {2020})}\BibitemShut {NoStop}%
\bibitem [{\citenamefont {Fulde}\ and\ \citenamefont
  {Ferrell}(1964)}]{PhysRev.135.A550}%
  \BibitemOpen
  \bibfield  {author} {\bibinfo {author} {\bibfnamefont {P.}~\bibnamefont
  {Fulde}}\ and\ \bibinfo {author} {\bibfnamefont {R.~A.}\ \bibnamefont
  {Ferrell}},\ }\bibfield  {title} {\bibinfo {title} {Superconductivity in a
  strong spin-exchange field},\ }\href
  {https://doi.org/10.1103/PhysRev.135.A550} {\bibfield  {journal} {\bibinfo
  {journal} {Phys. Rev.}\ }\textbf {\bibinfo {volume} {135}},\ \bibinfo {pages}
  {A550} (\bibinfo {year} {1964})}\BibitemShut {NoStop}%
\bibitem [{\citenamefont {Larkin}\ and\ \citenamefont
  {Ovchinnikov}(1964)}]{LO}%
  \BibitemOpen
  \bibfield  {author} {\bibinfo {author} {\bibfnamefont {A.~I.}\ \bibnamefont
  {Larkin}}\ and\ \bibinfo {author} {\bibfnamefont {Y.~N.}\ \bibnamefont
  {Ovchinnikov}},\ }\bibfield  {title} {\bibinfo {title} {Nonuniform state of
  superconductors},\ }\href@noop {} {\bibfield  {journal} {\bibinfo  {journal}
  {Zh. Eksp. Teor. Fiz. 47, 1136 (1964) [Sov. Phys. JETP 20, 762 (1965)]}\ }
  (\bibinfo {year} {1964})}\BibitemShut {NoStop}%
\bibitem [{\citenamefont {Kinnunen}\ \emph {et~al.}(2018)\citenamefont
  {Kinnunen}, \citenamefont {Baarsma}, \citenamefont {Martikainen},\ and\
  \citenamefont {T{\"o}rm{\"a}}}]{kinnunen2018fulde}%
  \BibitemOpen
  \bibfield  {author} {\bibinfo {author} {\bibfnamefont {J.~J.}\ \bibnamefont
  {Kinnunen}}, \bibinfo {author} {\bibfnamefont {J.~E.}\ \bibnamefont
  {Baarsma}}, \bibinfo {author} {\bibfnamefont {J.-P.}\ \bibnamefont
  {Martikainen}},\ and\ \bibinfo {author} {\bibfnamefont {P.}~\bibnamefont
  {T{\"o}rm{\"a}}},\ }\bibfield  {title} {\bibinfo {title} {The
  fulde--ferrell--larkin--ovchinnikov state for ultracold fermions in lattice
  and harmonic potentials: a review},\ }\href@noop {} {\bibfield  {journal}
  {\bibinfo  {journal} {Reports on Progress in Physics}\ }\textbf {\bibinfo
  {volume} {81}},\ \bibinfo {pages} {046401} (\bibinfo {year}
  {2018})}\BibitemShut {NoStop}%
\bibitem [{\citenamefont {Agterberg}\ and\ \citenamefont
  {Tsunetsugu}(2008)}]{PDW}%
  \BibitemOpen
  \bibfield  {author} {\bibinfo {author} {\bibfnamefont {D.~F.}\ \bibnamefont
  {Agterberg}}\ and\ \bibinfo {author} {\bibfnamefont {H.}~\bibnamefont
  {Tsunetsugu}},\ }\bibfield  {title} {\bibinfo {title} {Dislocations and
  vortices in pair-density-wave superconductors},\ }\href
  {https://doi.org/10.1038/nphys999} {\bibfield  {journal} {\bibinfo  {journal}
  {Nature Physics}\ }\textbf {\bibinfo {volume} {4}},\ \bibinfo {pages} {639}
  (\bibinfo {year} {2008})}\BibitemShut {NoStop}%
\bibitem [{\citenamefont {Berg}\ \emph
  {et~al.}(2009{\natexlab{a}})\citenamefont {Berg}, \citenamefont {Fradkin},\
  and\ \citenamefont {Kivelson}}]{PhysRevB.79.064515}%
  \BibitemOpen
  \bibfield  {author} {\bibinfo {author} {\bibfnamefont {E.}~\bibnamefont
  {Berg}}, \bibinfo {author} {\bibfnamefont {E.}~\bibnamefont {Fradkin}},\ and\
  \bibinfo {author} {\bibfnamefont {S.~A.}\ \bibnamefont {Kivelson}},\
  }\bibfield  {title} {\bibinfo {title} {Theory of the striped
  superconductor},\ }\href {https://doi.org/10.1103/PhysRevB.79.064515}
  {\bibfield  {journal} {\bibinfo  {journal} {Phys. Rev. B}\ }\textbf {\bibinfo
  {volume} {79}},\ \bibinfo {pages} {064515} (\bibinfo {year}
  {2009}{\natexlab{a}})}\BibitemShut {NoStop}%
\bibitem [{\citenamefont {Loder}\ \emph {et~al.}(2010)\citenamefont {Loder},
  \citenamefont {Kampf},\ and\ \citenamefont {Kopp}}]{PhysRevB.81.020511}%
  \BibitemOpen
  \bibfield  {author} {\bibinfo {author} {\bibfnamefont {F.}~\bibnamefont
  {Loder}}, \bibinfo {author} {\bibfnamefont {A.~P.}\ \bibnamefont {Kampf}},\
  and\ \bibinfo {author} {\bibfnamefont {T.}~\bibnamefont {Kopp}},\ }\bibfield
  {title} {\bibinfo {title} {Superconducting state with a finite-momentum
  pairing mechanism in zero external magnetic field},\ }\href
  {https://doi.org/10.1103/PhysRevB.81.020511} {\bibfield  {journal} {\bibinfo
  {journal} {Phys. Rev. B}\ }\textbf {\bibinfo {volume} {81}},\ \bibinfo
  {pages} {020511} (\bibinfo {year} {2010})}\BibitemShut {NoStop}%
\bibitem [{\citenamefont {Soto-Garrido}\ and\ \citenamefont
  {Fradkin}(2014)}]{PhysRevB.89.165126}%
  \BibitemOpen
  \bibfield  {author} {\bibinfo {author} {\bibfnamefont {R.}~\bibnamefont
  {Soto-Garrido}}\ and\ \bibinfo {author} {\bibfnamefont {E.}~\bibnamefont
  {Fradkin}},\ }\bibfield  {title} {\bibinfo {title} {Pair-density-wave
  superconducting states and electronic liquid-crystal phases},\ }\href
  {https://doi.org/10.1103/PhysRevB.89.165126} {\bibfield  {journal} {\bibinfo
  {journal} {Phys. Rev. B}\ }\textbf {\bibinfo {volume} {89}},\ \bibinfo
  {pages} {165126} (\bibinfo {year} {2014})}\BibitemShut {NoStop}%
\bibitem [{\citenamefont {Berg}\ \emph {et~al.}(2007)\citenamefont {Berg},
  \citenamefont {Fradkin}, \citenamefont {Kim}, \citenamefont {Kivelson},
  \citenamefont {Oganesyan}, \citenamefont {Tranquada},\ and\ \citenamefont
  {Zhang}}]{PhysRevLett.99.127003}%
  \BibitemOpen
  \bibfield  {author} {\bibinfo {author} {\bibfnamefont {E.}~\bibnamefont
  {Berg}}, \bibinfo {author} {\bibfnamefont {E.}~\bibnamefont {Fradkin}},
  \bibinfo {author} {\bibfnamefont {E.-A.}\ \bibnamefont {Kim}}, \bibinfo
  {author} {\bibfnamefont {S.~A.}\ \bibnamefont {Kivelson}}, \bibinfo {author}
  {\bibfnamefont {V.}~\bibnamefont {Oganesyan}}, \bibinfo {author}
  {\bibfnamefont {J.~M.}\ \bibnamefont {Tranquada}},\ and\ \bibinfo {author}
  {\bibfnamefont {S.~C.}\ \bibnamefont {Zhang}},\ }\bibfield  {title} {\bibinfo
  {title} {Dynamical layer decoupling in a stripe-ordered high-${T}_{c}$
  superconductor},\ }\href {https://doi.org/10.1103/PhysRevLett.99.127003}
  {\bibfield  {journal} {\bibinfo  {journal} {Phys. Rev. Lett.}\ }\textbf
  {\bibinfo {volume} {99}},\ \bibinfo {pages} {127003} (\bibinfo {year}
  {2007})}\BibitemShut {NoStop}%
\bibitem [{\citenamefont {Lee}(2014)}]{lee2014amperean}%
  \BibitemOpen
  \bibfield  {author} {\bibinfo {author} {\bibfnamefont {P.~A.}\ \bibnamefont
  {Lee}},\ }\bibfield  {title} {\bibinfo {title} {Amperean pairing and the
  pseudogap phase of cuprate superconductors},\ }\href@noop {} {\bibfield
  {journal} {\bibinfo  {journal} {Physical Review X}\ }\textbf {\bibinfo
  {volume} {4}},\ \bibinfo {pages} {031017} (\bibinfo {year}
  {2014})}\BibitemShut {NoStop}%
\bibitem [{\citenamefont {Berg}\ \emph
  {et~al.}(2009{\natexlab{b}})\citenamefont {Berg}, \citenamefont {Fradkin},\
  and\ \citenamefont {Kivelson}}]{berg2009charge}%
  \BibitemOpen
  \bibfield  {author} {\bibinfo {author} {\bibfnamefont {E.}~\bibnamefont
  {Berg}}, \bibinfo {author} {\bibfnamefont {E.}~\bibnamefont {Fradkin}},\ and\
  \bibinfo {author} {\bibfnamefont {S.~A.}\ \bibnamefont {Kivelson}},\
  }\bibfield  {title} {\bibinfo {title} {Charge-4 e superconductivity from
  pair-density-wave order in certain high-temperature superconductors},\
  }\href@noop {} {\bibfield  {journal} {\bibinfo  {journal} {Nature Physics}\
  }\textbf {\bibinfo {volume} {5}},\ \bibinfo {pages} {830} (\bibinfo {year}
  {2009}{\natexlab{b}})}\BibitemShut {NoStop}%
\bibitem [{\citenamefont {Wang}\ \emph {et~al.}(2018)\citenamefont {Wang},
  \citenamefont {Edkins}, \citenamefont {Hamidian}, \citenamefont {Davis},
  \citenamefont {Fradkin},\ and\ \citenamefont {Kivelson}}]{wang2018pair}%
  \BibitemOpen
  \bibfield  {author} {\bibinfo {author} {\bibfnamefont {Y.}~\bibnamefont
  {Wang}}, \bibinfo {author} {\bibfnamefont {S.~D.}\ \bibnamefont {Edkins}},
  \bibinfo {author} {\bibfnamefont {M.~H.}\ \bibnamefont {Hamidian}}, \bibinfo
  {author} {\bibfnamefont {J.~S.}\ \bibnamefont {Davis}}, \bibinfo {author}
  {\bibfnamefont {E.}~\bibnamefont {Fradkin}},\ and\ \bibinfo {author}
  {\bibfnamefont {S.~A.}\ \bibnamefont {Kivelson}},\ }\bibfield  {title}
  {\bibinfo {title} {Pair density waves in superconducting vortex halos},\
  }\href@noop {} {\bibfield  {journal} {\bibinfo  {journal} {Physical Review
  B}\ }\textbf {\bibinfo {volume} {97}},\ \bibinfo {pages} {174510} (\bibinfo
  {year} {2018})}\BibitemShut {NoStop}%
\bibitem [{\citenamefont {Zhang}\ \emph
  {et~al.}(2024{\natexlab{a}})\citenamefont {Zhang}, \citenamefont {Hu},\ and\
  \citenamefont {Neupert}}]{altermagnets}%
  \BibitemOpen
  \bibfield  {author} {\bibinfo {author} {\bibfnamefont {S.-B.}\ \bibnamefont
  {Zhang}}, \bibinfo {author} {\bibfnamefont {L.-H.}\ \bibnamefont {Hu}},\ and\
  \bibinfo {author} {\bibfnamefont {T.}~\bibnamefont {Neupert}},\ }\bibfield
  {title} {\bibinfo {title} {Finite-momentum cooper pairing in proximitized
  altermagnets},\ }\bibfield  {journal} {\bibinfo  {journal} {Nature
  Communications}\ }\textbf {\bibinfo {volume} {15}},\ \href
  {https://doi.org/10.1038/s41467-024-45951-3} {10.1038/s41467-024-45951-3}
  (\bibinfo {year} {2024}{\natexlab{a}})\BibitemShut {NoStop}%
\bibitem [{\citenamefont {Chakraborty}\ and\ \citenamefont
  {Black-Schaffer}(2024{\natexlab{a}})}]{PhysRevB.110.L060508}%
  \BibitemOpen
  \bibfield  {author} {\bibinfo {author} {\bibfnamefont {D.}~\bibnamefont
  {Chakraborty}}\ and\ \bibinfo {author} {\bibfnamefont {A.~M.}\ \bibnamefont
  {Black-Schaffer}},\ }\bibfield  {title} {\bibinfo {title} {Zero-field
  finite-momentum and field-induced superconductivity in altermagnets},\ }\href
  {https://doi.org/10.1103/PhysRevB.110.L060508} {\bibfield  {journal}
  {\bibinfo  {journal} {Phys. Rev. B}\ }\textbf {\bibinfo {volume} {110}},\
  \bibinfo {pages} {L060508} (\bibinfo {year}
  {2024}{\natexlab{a}})}\BibitemShut {NoStop}%
\bibitem [{\citenamefont {Sim}\ and\ \citenamefont
  {Knolle}(2024)}]{sim2024pair}%
  \BibitemOpen
  \bibfield  {author} {\bibinfo {author} {\bibfnamefont {G.}~\bibnamefont
  {Sim}}\ and\ \bibinfo {author} {\bibfnamefont {J.}~\bibnamefont {Knolle}},\
  }\bibfield  {title} {\bibinfo {title} {Pair density waves and supercurrent
  diode effect in altermagnets},\ }\href@noop {} {\bibfield  {journal}
  {\bibinfo  {journal} {arXiv preprint arXiv:2407.01513}\ } (\bibinfo {year}
  {2024})}\BibitemShut {NoStop}%
\bibitem [{\citenamefont {Chakraborty}\ and\ \citenamefont
  {Black-Schaffer}(2024{\natexlab{b}})}]{chakraborty2024perfect}%
  \BibitemOpen
  \bibfield  {author} {\bibinfo {author} {\bibfnamefont {D.}~\bibnamefont
  {Chakraborty}}\ and\ \bibinfo {author} {\bibfnamefont {A.~M.}\ \bibnamefont
  {Black-Schaffer}},\ }\bibfield  {title} {\bibinfo {title} {Perfect
  superconducting diode effect in altermagnets},\ }\href@noop {} {\bibfield
  {journal} {\bibinfo  {journal} {arXiv preprint arXiv:2408.07747}\ } (\bibinfo
  {year} {2024}{\natexlab{b}})}\BibitemShut {NoStop}%
\bibitem [{\citenamefont {Sun}\ \emph {et~al.}(2024)\citenamefont {Sun},
  \citenamefont {Yu}, \citenamefont {Chen}, \citenamefont {Hu},\ and\
  \citenamefont
  {Law}}]{sun2024flatbandfuldeferrelllarkinovchinnikovstatequantum}%
  \BibitemOpen
  \bibfield  {author} {\bibinfo {author} {\bibfnamefont {Z.-T.}\ \bibnamefont
  {Sun}}, \bibinfo {author} {\bibfnamefont {R.-P.}\ \bibnamefont {Yu}},
  \bibinfo {author} {\bibfnamefont {S.~A.}\ \bibnamefont {Chen}}, \bibinfo
  {author} {\bibfnamefont {J.-X.}\ \bibnamefont {Hu}},\ and\ \bibinfo {author}
  {\bibfnamefont {K.~T.}\ \bibnamefont {Law}},\ }\bibfield  {title} {\bibinfo
  {title} {Flat-band fulde-ferrell-larkin-ovchinnikov state from quantum
  geometry discrepancy},\ }\href@noop {} {\bibfield  {journal} {\bibinfo
  {journal} {arXiv preprint arXiv:2408.00548}\ } (\bibinfo {year}
  {2024})}\BibitemShut {NoStop}%
\bibitem [{\citenamefont {Takada}(1970{\natexlab{a}})}]{10.1143/PTP.43.27}%
  \BibitemOpen
  \bibfield  {author} {\bibinfo {author} {\bibfnamefont {S.}~\bibnamefont
  {Takada}},\ }\bibfield  {title} {\bibinfo {title} {Superconductivity in a
  molecular field. ii: Stability of fulde-ferrel phase},\ }\href
  {https://doi.org/10.1143/PTP.43.27} {\bibfield  {journal} {\bibinfo
  {journal} {Progress of Theoretical Physics}\ }\textbf {\bibinfo {volume}
  {43}},\ \bibinfo {pages} {27} (\bibinfo {year}
  {1970}{\natexlab{a}})}\BibitemShut {NoStop}%
\bibitem [{\citenamefont {Wang}\ \emph {et~al.}(2007)\citenamefont {Wang},
  \citenamefont {Hu},\ and\ \citenamefont {Ting}}]{PhysRevB.75.184515}%
  \BibitemOpen
  \bibfield  {author} {\bibinfo {author} {\bibfnamefont {Q.}~\bibnamefont
  {Wang}}, \bibinfo {author} {\bibfnamefont {C.-R.}\ \bibnamefont {Hu}},\ and\
  \bibinfo {author} {\bibfnamefont {C.-S.}\ \bibnamefont {Ting}},\ }\bibfield
  {title} {\bibinfo {title} {Impurity-induced configuration-transition in the
  fulde-ferrell-larkin-ovchinnikov state of a $d$-wave superconductor},\ }\href
  {https://doi.org/10.1103/PhysRevB.75.184515} {\bibfield  {journal} {\bibinfo
  {journal} {Phys. Rev. B}\ }\textbf {\bibinfo {volume} {75}},\ \bibinfo
  {pages} {184515} (\bibinfo {year} {2007})}\BibitemShut {NoStop}%
\bibitem [{\citenamefont {Gruenberg}\ and\ \citenamefont
  {Gunther}(1966)}]{PhysRevLett.16.996}%
  \BibitemOpen
  \bibfield  {author} {\bibinfo {author} {\bibfnamefont {L.~W.}\ \bibnamefont
  {Gruenberg}}\ and\ \bibinfo {author} {\bibfnamefont {L.}~\bibnamefont
  {Gunther}},\ }\bibfield  {title} {\bibinfo {title} {Fulde-ferrell effect in
  type-ii superconductors},\ }\href
  {https://doi.org/10.1103/PhysRevLett.16.996} {\bibfield  {journal} {\bibinfo
  {journal} {Phys. Rev. Lett.}\ }\textbf {\bibinfo {volume} {16}},\ \bibinfo
  {pages} {996} (\bibinfo {year} {1966})}\BibitemShut {NoStop}%
\bibitem [{\citenamefont {Adachi}\ and\ \citenamefont
  {Ikeda}(2003)}]{PhysRevB.68.184510}%
  \BibitemOpen
  \bibfield  {author} {\bibinfo {author} {\bibfnamefont {H.}~\bibnamefont
  {Adachi}}\ and\ \bibinfo {author} {\bibfnamefont {R.}~\bibnamefont {Ikeda}},\
  }\bibfield  {title} {\bibinfo {title} {Effects of pauli paramagnetism on the
  superconducting vortex phase diagram in strong fields},\ }\href
  {https://doi.org/10.1103/PhysRevB.68.184510} {\bibfield  {journal} {\bibinfo
  {journal} {Phys. Rev. B}\ }\textbf {\bibinfo {volume} {68}},\ \bibinfo
  {pages} {184510} (\bibinfo {year} {2003})}\BibitemShut {NoStop}%
\bibitem [{\citenamefont {Chen}\ and\ \citenamefont
  {Huang}(2023)}]{chen2023pair}%
  \BibitemOpen
  \bibfield  {author} {\bibinfo {author} {\bibfnamefont {W.}~\bibnamefont
  {Chen}}\ and\ \bibinfo {author} {\bibfnamefont {W.}~\bibnamefont {Huang}},\
  }\bibfield  {title} {\bibinfo {title} {Pair density wave facilitated by bloch
  quantum geometry in nearly flat band multiorbital superconductors},\
  }\href@noop {} {\bibfield  {journal} {\bibinfo  {journal} {Science China
  Physics, Mechanics \& Astronomy}\ }\textbf {\bibinfo {volume} {66}},\
  \bibinfo {pages} {287212} (\bibinfo {year} {2023})}\BibitemShut {NoStop}%
\bibitem [{\citenamefont {Jiang}\ and\ \citenamefont
  {Barlas}(2023)}]{jiang2023pair}%
  \BibitemOpen
  \bibfield  {author} {\bibinfo {author} {\bibfnamefont {G.}~\bibnamefont
  {Jiang}}\ and\ \bibinfo {author} {\bibfnamefont {Y.}~\bibnamefont {Barlas}},\
  }\bibfield  {title} {\bibinfo {title} {Pair density waves from local band
  geometry},\ }\href@noop {} {\bibfield  {journal} {\bibinfo  {journal}
  {Physical Review Letters}\ }\textbf {\bibinfo {volume} {131}},\ \bibinfo
  {pages} {016002} (\bibinfo {year} {2023})}\BibitemShut {NoStop}%
\bibitem [{\citenamefont {Wang}\ and\ \citenamefont
  {Huang}(2024)}]{wang2024quantum}%
  \BibitemOpen
  \bibfield  {author} {\bibinfo {author} {\bibfnamefont {H.-X.}\ \bibnamefont
  {Wang}}\ and\ \bibinfo {author} {\bibfnamefont {W.}~\bibnamefont {Huang}},\
  }\bibfield  {title} {\bibinfo {title} {Quantum-geometry-facilitated pair
  density wave order: Density matrix renormalization group study},\ }\href@noop
  {} {\bibfield  {journal} {\bibinfo  {journal} {arXiv preprint
  arXiv:2406.17187}\ } (\bibinfo {year} {2024})}\BibitemShut {NoStop}%
\bibitem [{\citenamefont {Chen}\ \emph {et~al.}(2004)\citenamefont {Chen},
  \citenamefont {Vafek}, \citenamefont {Yazdani},\ and\ \citenamefont
  {Zhang}}]{PhysRevLett.93.187002}%
  \BibitemOpen
  \bibfield  {author} {\bibinfo {author} {\bibfnamefont {H.-D.}\ \bibnamefont
  {Chen}}, \bibinfo {author} {\bibfnamefont {O.}~\bibnamefont {Vafek}},
  \bibinfo {author} {\bibfnamefont {A.}~\bibnamefont {Yazdani}},\ and\ \bibinfo
  {author} {\bibfnamefont {S.-C.}\ \bibnamefont {Zhang}},\ }\bibfield  {title}
  {\bibinfo {title} {Pair density wave in the pseudogap state of high
  temperature superconductors},\ }\href
  {https://doi.org/10.1103/PhysRevLett.93.187002} {\bibfield  {journal}
  {\bibinfo  {journal} {Phys. Rev. Lett.}\ }\textbf {\bibinfo {volume} {93}},\
  \bibinfo {pages} {187002} (\bibinfo {year} {2004})}\BibitemShut {NoStop}%
\bibitem [{\citenamefont {Agterberg}\ and\ \citenamefont
  {Kaur}(2007)}]{PhysRevB.75.064511}%
  \BibitemOpen
  \bibfield  {author} {\bibinfo {author} {\bibfnamefont {D.~F.}\ \bibnamefont
  {Agterberg}}\ and\ \bibinfo {author} {\bibfnamefont {R.~P.}\ \bibnamefont
  {Kaur}},\ }\bibfield  {title} {\bibinfo {title} {Magnetic-field-induced
  helical and stripe phases in rashba superconductors},\ }\href
  {https://doi.org/10.1103/PhysRevB.75.064511} {\bibfield  {journal} {\bibinfo
  {journal} {Phys. Rev. B}\ }\textbf {\bibinfo {volume} {75}},\ \bibinfo
  {pages} {064511} (\bibinfo {year} {2007})}\BibitemShut {NoStop}%
\bibitem [{\citenamefont {Dimitrova}\ and\ \citenamefont
  {Feigel'man}(2007)}]{PhysRevB.76.014522}%
  \BibitemOpen
  \bibfield  {author} {\bibinfo {author} {\bibfnamefont {O.}~\bibnamefont
  {Dimitrova}}\ and\ \bibinfo {author} {\bibfnamefont {M.~V.}\ \bibnamefont
  {Feigel'man}},\ }\bibfield  {title} {\bibinfo {title} {Theory of a
  two-dimensional superconductor with broken inversion symmetry},\ }\href
  {https://doi.org/10.1103/PhysRevB.76.014522} {\bibfield  {journal} {\bibinfo
  {journal} {Phys. Rev. B}\ }\textbf {\bibinfo {volume} {76}},\ \bibinfo
  {pages} {014522} (\bibinfo {year} {2007})}\BibitemShut {NoStop}%
\bibitem [{\citenamefont {Yuan}\ and\ \citenamefont
  {Fu}(2021)}]{Topologicalmetalsandfinite-momentumsuperconductors}%
  \BibitemOpen
  \bibfield  {author} {\bibinfo {author} {\bibfnamefont {N.~F.~Q.}\
  \bibnamefont {Yuan}}\ and\ \bibinfo {author} {\bibfnamefont {L.}~\bibnamefont
  {Fu}},\ }\bibfield  {title} {\bibinfo {title} {Topological metals and
  finite-momentum superconductors},\ }\href@noop {} {\bibfield  {journal}
  {\bibinfo  {journal} {Proc Natl Acad Sci U S A}\ }\textbf {\bibinfo {volume}
  {118}} (\bibinfo {year} {2021})}\BibitemShut {NoStop}%
\bibitem [{\citenamefont {Yuan}\ and\ \citenamefont
  {Fu}(2022)}]{Supercurrentdiodeeffectandfinite-momentumsuperconductors}%
  \BibitemOpen
  \bibfield  {author} {\bibinfo {author} {\bibfnamefont {N.~F.~Q.}\
  \bibnamefont {Yuan}}\ and\ \bibinfo {author} {\bibfnamefont {L.}~\bibnamefont
  {Fu}},\ }\bibfield  {title} {\bibinfo {title} {Supercurrent diode effect and
  finite-momentum superconductors},\ }\href@noop {} {\bibfield  {journal}
  {\bibinfo  {journal} {Proc Natl Acad Sci U S A}\ }\textbf {\bibinfo {volume}
  {119}},\ \bibinfo {pages} {e2119548119} (\bibinfo {year} {2022})}\BibitemShut
  {NoStop}%
\bibitem [{\citenamefont {Rashba}(1960)}]{EIRashba}%
  \BibitemOpen
  \bibfield  {author} {\bibinfo {author} {\bibfnamefont {E.~I.}\ \bibnamefont
  {Rashba}},\ }\bibfield  {title} {\bibinfo {title} {Properties of
  semiconductors with an extremum loop .1. cyclotron and combinational
  resonance in a magnetic field perpendicular to the plane of the loop},\
  }\href@noop {} {\bibfield  {journal} {\bibinfo  {journal} {Sov. Phys. Solid.
  State}\ }\textbf {\bibinfo {volume} {2}},\ \bibinfo {pages} {1109} (\bibinfo
  {year} {1960})}\BibitemShut {NoStop}%
\bibitem [{\citenamefont {Bychkov}\ and\ \citenamefont
  {Rashba}(1984)}]{PismaZhETF.39.66}%
  \BibitemOpen
  \bibfield  {author} {\bibinfo {author} {\bibfnamefont {Y.~A.}\ \bibnamefont
  {Bychkov}}\ and\ \bibinfo {author} {\bibfnamefont {E.~I.}\ \bibnamefont
  {Rashba}},\ }\bibfield  {title} {\bibinfo {title} {Properties of a 2d
  electron-gas with lifted spectral degeneracy},\ }\href {<Go to
  ISI>://WOS:A1984SY04100009} {\bibfield  {journal} {\bibinfo  {journal} {Jetp
  Letters}\ }\textbf {\bibinfo {volume} {39}},\ \bibinfo {pages} {78} (\bibinfo
  {year} {1984})}\BibitemShut {NoStop}%
\bibitem [{\citenamefont {Winkler}\ and\ \citenamefont
  {R\"ossler}(1993)}]{PhysRevB.48.8918}%
  \BibitemOpen
  \bibfield  {author} {\bibinfo {author} {\bibfnamefont {R.}~\bibnamefont
  {Winkler}}\ and\ \bibinfo {author} {\bibfnamefont {U.}~\bibnamefont
  {R\"ossler}},\ }\bibfield  {title} {\bibinfo {title} {General approach to the
  envelope-function approximation based on a quadrature method},\ }\href
  {https://doi.org/10.1103/PhysRevB.48.8918} {\bibfield  {journal} {\bibinfo
  {journal} {Phys. Rev. B}\ }\textbf {\bibinfo {volume} {48}},\ \bibinfo
  {pages} {8918} (\bibinfo {year} {1993})}\BibitemShut {NoStop}%
\bibitem [{\citenamefont {Winkler}(2000)}]{PhysRevB.62.4245}%
  \BibitemOpen
  \bibfield  {author} {\bibinfo {author} {\bibfnamefont {R.}~\bibnamefont
  {Winkler}},\ }\bibfield  {title} {\bibinfo {title} {Rashba spin splitting in
  two-dimensional electron and hole systems},\ }\href
  {https://doi.org/10.1103/PhysRevB.62.4245} {\bibfield  {journal} {\bibinfo
  {journal} {Phys. Rev. B}\ }\textbf {\bibinfo {volume} {62}},\ \bibinfo
  {pages} {4245} (\bibinfo {year} {2000})}\BibitemShut {NoStop}%
\bibitem [{\citenamefont {Gor'kov}\ and\ \citenamefont
  {Rashba}(2001)}]{PhysRevLett.87.037004}%
  \BibitemOpen
  \bibfield  {author} {\bibinfo {author} {\bibfnamefont {L.~P.}\ \bibnamefont
  {Gor'kov}}\ and\ \bibinfo {author} {\bibfnamefont {E.~I.}\ \bibnamefont
  {Rashba}},\ }\bibfield  {title} {\bibinfo {title} {Superconducting 2d system
  with lifted spin degeneracy: Mixed singlet-triplet state},\ }\href
  {https://doi.org/10.1103/PhysRevLett.87.037004} {\bibfield  {journal}
  {\bibinfo  {journal} {Phys. Rev. Lett.}\ }\textbf {\bibinfo {volume} {87}},\
  \bibinfo {pages} {037004} (\bibinfo {year} {2001})}\BibitemShut {NoStop}%
\bibitem [{\citenamefont {Frigeri}\ \emph {et~al.}(2004)\citenamefont
  {Frigeri}, \citenamefont {Agterberg}, \citenamefont {Koga},\ and\
  \citenamefont {Sigrist}}]{PhysRevLett.92.097001}%
  \BibitemOpen
  \bibfield  {author} {\bibinfo {author} {\bibfnamefont {P.~A.}\ \bibnamefont
  {Frigeri}}, \bibinfo {author} {\bibfnamefont {D.~F.}\ \bibnamefont
  {Agterberg}}, \bibinfo {author} {\bibfnamefont {A.}~\bibnamefont {Koga}},\
  and\ \bibinfo {author} {\bibfnamefont {M.}~\bibnamefont {Sigrist}},\
  }\bibfield  {title} {\bibinfo {title} {Superconductivity without inversion
  symmetry: Mnsi versus
  ${\mathrm{c}\mathrm{e}\mathrm{p}\mathrm{t}}_{3}\mathrm{S}\mathrm{i}$},\
  }\href {https://doi.org/10.1103/PhysRevLett.92.097001} {\bibfield  {journal}
  {\bibinfo  {journal} {Phys. Rev. Lett.}\ }\textbf {\bibinfo {volume} {92}},\
  \bibinfo {pages} {097001} (\bibinfo {year} {2004})}\BibitemShut {NoStop}%
\bibitem [{\citenamefont {Huang}\ \emph {et~al.}(2024)\citenamefont {Huang},
  \citenamefont {Xiao}, \citenamefont {Song},\ and\ \citenamefont
  {Hao}}]{PhysRevB.109.195419}%
  \BibitemOpen
  \bibfield  {author} {\bibinfo {author} {\bibfnamefont {X.}~\bibnamefont
  {Huang}}, \bibinfo {author} {\bibfnamefont {Y.}~\bibnamefont {Xiao}},
  \bibinfo {author} {\bibfnamefont {R.}~\bibnamefont {Song}},\ and\ \bibinfo
  {author} {\bibfnamefont {N.}~\bibnamefont {Hao}},\ }\bibfield  {title}
  {\bibinfo {title} {Generic model with unconventional rashba bands and giant
  spin galvanic effect},\ }\href {https://doi.org/10.1103/PhysRevB.109.195419}
  {\bibfield  {journal} {\bibinfo  {journal} {Phys. Rev. B}\ }\textbf {\bibinfo
  {volume} {109}},\ \bibinfo {pages} {195419} (\bibinfo {year}
  {2024})}\BibitemShut {NoStop}%
\bibitem [{\citenamefont {Song}\ \emph {et~al.}(2021)\citenamefont {Song},
  \citenamefont {Hao},\ and\ \citenamefont {Zhang}}]{song2021giant}%
  \BibitemOpen
  \bibfield  {author} {\bibinfo {author} {\bibfnamefont {R.}~\bibnamefont
  {Song}}, \bibinfo {author} {\bibfnamefont {N.}~\bibnamefont {Hao}},\ and\
  \bibinfo {author} {\bibfnamefont {P.}~\bibnamefont {Zhang}},\ }\bibfield
  {title} {\bibinfo {title} {Giant inverse rashba-edelstein effect: Application
  to monolayer osbi 2},\ }\href@noop {} {\bibfield  {journal} {\bibinfo
  {journal} {Physical Review B}\ }\textbf {\bibinfo {volume} {104}},\ \bibinfo
  {pages} {115433} (\bibinfo {year} {2021})}\BibitemShut {NoStop}%
\bibitem [{\citenamefont {Wang}\ \emph {et~al.}(2024)\citenamefont {Wang},
  \citenamefont {Li}, \citenamefont {Huang}, \citenamefont {Wang},
  \citenamefont {Song},\ and\ \citenamefont
  {Hao}}]{wang2024superconductivitytwodimensionalsystemsunconventional}%
  \BibitemOpen
  \bibfield  {author} {\bibinfo {author} {\bibfnamefont {R.}~\bibnamefont
  {Wang}}, \bibinfo {author} {\bibfnamefont {J.}~\bibnamefont {Li}}, \bibinfo
  {author} {\bibfnamefont {X.}~\bibnamefont {Huang}}, \bibinfo {author}
  {\bibfnamefont {L.}~\bibnamefont {Wang}}, \bibinfo {author} {\bibfnamefont
  {R.}~\bibnamefont {Song}},\ and\ \bibinfo {author} {\bibfnamefont
  {N.}~\bibnamefont {Hao}},\ }\bibfield  {title} {\bibinfo {title}
  {Superconductivity in two-dimensional systems with unconventional rashba
  bands},\ }\href {https://doi.org/10.1103/PhysRevB.110.134517} {\bibfield
  {journal} {\bibinfo  {journal} {Phys. Rev. B}\ }\textbf {\bibinfo {volume}
  {110}},\ \bibinfo {pages} {134517} (\bibinfo {year} {2024})}\BibitemShut
  {NoStop}%
\bibitem [{\citenamefont {Bhattacharya}\ and\ \citenamefont
  {Black-Schaffer}(2024)}]{bhattacharya2024electricfieldinducedsecondorder}%
  \BibitemOpen
  \bibfield  {author} {\bibinfo {author} {\bibfnamefont {A.}~\bibnamefont
  {Bhattacharya}}\ and\ \bibinfo {author} {\bibfnamefont {A.~M.}\ \bibnamefont
  {Black-Schaffer}},\ }\bibfield  {title} {\bibinfo {title} {Electric field
  induced second-order anomalous hall transport in unconventional rashba
  system},\ }\href@noop {} {\bibfield  {journal} {\bibinfo  {journal} {arXiv:
  2408.15840}\ } (\bibinfo {year} {2024})}\BibitemShut {NoStop}%
\bibitem [{\citenamefont {Gao}\ \emph {et~al.}(2018)\citenamefont {Gao},
  \citenamefont {Zhu}, \citenamefont {Zheng}, \citenamefont {Wu}, \citenamefont
  {Zhang}, \citenamefont {Xi}, \citenamefont {Zhang}, \citenamefont {Zhang},
  \citenamefont {Hao}, \citenamefont {Ning} \emph {et~al.}}]{gao2018possible}%
  \BibitemOpen
  \bibfield  {author} {\bibinfo {author} {\bibfnamefont {W.}~\bibnamefont
  {Gao}}, \bibinfo {author} {\bibfnamefont {X.}~\bibnamefont {Zhu}}, \bibinfo
  {author} {\bibfnamefont {F.}~\bibnamefont {Zheng}}, \bibinfo {author}
  {\bibfnamefont {M.}~\bibnamefont {Wu}}, \bibinfo {author} {\bibfnamefont
  {J.}~\bibnamefont {Zhang}}, \bibinfo {author} {\bibfnamefont
  {C.}~\bibnamefont {Xi}}, \bibinfo {author} {\bibfnamefont {P.}~\bibnamefont
  {Zhang}}, \bibinfo {author} {\bibfnamefont {Y.}~\bibnamefont {Zhang}},
  \bibinfo {author} {\bibfnamefont {N.}~\bibnamefont {Hao}}, \bibinfo {author}
  {\bibfnamefont {W.}~\bibnamefont {Ning}}, \emph {et~al.},\ }\bibfield
  {title} {\bibinfo {title} {A possible candidate for triply degenerate point
  fermions in trigonal layered ptbi2},\ }\href@noop {} {\bibfield  {journal}
  {\bibinfo  {journal} {Nature communications}\ }\textbf {\bibinfo {volume}
  {9}},\ \bibinfo {pages} {3249} (\bibinfo {year} {2018})}\BibitemShut
  {NoStop}%
\bibitem [{\citenamefont {Gor'kov}\ and\ \citenamefont
  {P.}(1984)}]{AnunusualsuperconductivityinUBe13}%
  \BibitemOpen
  \bibfield  {author} {\bibinfo {author} {\bibfnamefont {G.~E.~V.}\
  \bibnamefont {Gor'kov}}\ and\ \bibinfo {author} {\bibfnamefont
  {L.}~\bibnamefont {P.}},\ }\bibfield  {title} {\bibinfo {title} {An unusual
  superconductivity in ube13},\ }\href@noop {} {\bibfield  {journal} {\bibinfo
  {journal} {JETP}\ } (\bibinfo {year} {1984})}\BibitemShut {NoStop}%
\bibitem [{\citenamefont {Gor'kov}\ and\ \citenamefont
  {P.}(1985)}]{Superconductingclassesinheavy-fermionsystems}%
  \BibitemOpen
  \bibfield  {author} {\bibinfo {author} {\bibfnamefont {G.~E.~V.}\
  \bibnamefont {Gor'kov}}\ and\ \bibinfo {author} {\bibfnamefont
  {L.}~\bibnamefont {P.}},\ }\bibfield  {title} {\bibinfo {title}
  {Superconducting classes in heavy-fermion systems},\ }\href@noop {}
  {\bibfield  {journal} {\bibinfo  {journal} {Soviet Physics JETP}\ } (\bibinfo
  {year} {1985})}\BibitemShut {NoStop}%
\bibitem [{\citenamefont {Ueda}\ and\ \citenamefont
  {Rice}(1985)}]{P-WaveSuperconductivityinCubicMetals}%
  \BibitemOpen
  \bibfield  {author} {\bibinfo {author} {\bibfnamefont {K.}~\bibnamefont
  {Ueda}}\ and\ \bibinfo {author} {\bibfnamefont {T.~M.}\ \bibnamefont
  {Rice}},\ }\bibfield  {title} {\bibinfo {title} {P-wave superconductivity in
  cubic metals},\ }\href {https://doi.org/DOI 10.1103/PhysRevB.31.7114}
  {\bibfield  {journal} {\bibinfo  {journal} {Physical Review B}\ }\textbf
  {\bibinfo {volume} {31}},\ \bibinfo {pages} {7114} (\bibinfo {year}
  {1985})}\BibitemShut {NoStop}%
\bibitem [{\citenamefont {Sigrist}\ and\ \citenamefont
  {Ueda}(1991)}]{RevModPhys.63.239}%
  \BibitemOpen
  \bibfield  {author} {\bibinfo {author} {\bibfnamefont {M.}~\bibnamefont
  {Sigrist}}\ and\ \bibinfo {author} {\bibfnamefont {K.}~\bibnamefont {Ueda}},\
  }\bibfield  {title} {\bibinfo {title} {Phenomenological theory of
  unconventional superconductivity},\ }\href
  {https://doi.org/10.1103/RevModPhys.63.239} {\bibfield  {journal} {\bibinfo
  {journal} {Rev. Mod. Phys.}\ }\textbf {\bibinfo {volume} {63}},\ \bibinfo
  {pages} {239} (\bibinfo {year} {1991})}\BibitemShut {NoStop}%
\bibitem [{\citenamefont {Wen}\ \emph {et~al.}(2022)\citenamefont {Wen},
  \citenamefont {Zhu}, \citenamefont {Xiao}, \citenamefont {Hao}, \citenamefont
  {Mondaini}, \citenamefont {Guo},\ and\ \citenamefont
  {Feng}}]{wen2022superconducting}%
  \BibitemOpen
  \bibfield  {author} {\bibinfo {author} {\bibfnamefont {C.}~\bibnamefont
  {Wen}}, \bibinfo {author} {\bibfnamefont {X.}~\bibnamefont {Zhu}}, \bibinfo
  {author} {\bibfnamefont {Z.}~\bibnamefont {Xiao}}, \bibinfo {author}
  {\bibfnamefont {N.}~\bibnamefont {Hao}}, \bibinfo {author} {\bibfnamefont
  {R.}~\bibnamefont {Mondaini}}, \bibinfo {author} {\bibfnamefont
  {H.}~\bibnamefont {Guo}},\ and\ \bibinfo {author} {\bibfnamefont
  {S.}~\bibnamefont {Feng}},\ }\bibfield  {title} {\bibinfo {title}
  {Superconducting pairing symmetry in the kagome-lattice hubbard model},\
  }\href@noop {} {\bibfield  {journal} {\bibinfo  {journal} {Physical Review
  B}\ }\textbf {\bibinfo {volume} {105}},\ \bibinfo {pages} {075118} (\bibinfo
  {year} {2022})}\BibitemShut {NoStop}%
\bibitem [{\citenamefont {Yang}\ \emph {et~al.}(2020)\citenamefont {Yang},
  \citenamefont {Mo}, \citenamefont {Fu}, \citenamefont {Yang}, \citenamefont
  {Zheng}, \citenamefont {Wang}, \citenamefont {Liu}, \citenamefont {Hao},\
  and\ \citenamefont {Zhang}}]{yang2020soft}%
  \BibitemOpen
  \bibfield  {author} {\bibinfo {author} {\bibfnamefont {W.}~\bibnamefont
  {Yang}}, \bibinfo {author} {\bibfnamefont {C.-J.}\ \bibnamefont {Mo}},
  \bibinfo {author} {\bibfnamefont {S.-B.}\ \bibnamefont {Fu}}, \bibinfo
  {author} {\bibfnamefont {Y.}~\bibnamefont {Yang}}, \bibinfo {author}
  {\bibfnamefont {F.-W.}\ \bibnamefont {Zheng}}, \bibinfo {author}
  {\bibfnamefont {X.-H.}\ \bibnamefont {Wang}}, \bibinfo {author}
  {\bibfnamefont {Y.-A.}\ \bibnamefont {Liu}}, \bibinfo {author} {\bibfnamefont
  {N.}~\bibnamefont {Hao}},\ and\ \bibinfo {author} {\bibfnamefont
  {P.}~\bibnamefont {Zhang}},\ }\bibfield  {title} {\bibinfo {title}
  {Soft-mode-phonon-mediated unconventional superconductivity in monolayer 1
  t$^{\prime}$-wte 2},\ }\href@noop {} {\bibfield  {journal} {\bibinfo
  {journal} {Physical Review Letters}\ }\textbf {\bibinfo {volume} {125}},\
  \bibinfo {pages} {237006} (\bibinfo {year} {2020})}\BibitemShut {NoStop}%
\bibitem [{sm()}]{sm}%
  \BibitemOpen
  \bibfield  {title} {\bibinfo {title} {See Supplemental Material
at [URL will be inserted by publisher] for the details of the derivation of the linearized gap equations
near T$_c$ for the intrinsic superconductivity, the situation with impurities, the proximity
effect, the derivation of Cooper-pair propagators, and the derivation of the Joesphson supercurrent}}\href@noop
  {} {\ }\BibitemShut {NoStop}%
\bibitem [{\citenamefont {Sheehy}\ and\ \citenamefont
  {Radzihovsky}(2006)}]{PhysRevLett.96.060401}%
  \BibitemOpen
  \bibfield  {author} {\bibinfo {author} {\bibfnamefont {D.~E.}\ \bibnamefont
  {Sheehy}}\ and\ \bibinfo {author} {\bibfnamefont {L.}~\bibnamefont
  {Radzihovsky}},\ }\bibfield  {title} {\bibinfo {title} {Bec-bcs crossover in
  ``magnetized'' feshbach-resonantly paired superfluids},\ }\href
  {https://doi.org/10.1103/PhysRevLett.96.060401} {\bibfield  {journal}
  {\bibinfo  {journal} {Phys. Rev. Lett.}\ }\textbf {\bibinfo {volume} {96}},\
  \bibinfo {pages} {060401} (\bibinfo {year} {2006})}\BibitemShut {NoStop}%
\bibitem [{\citenamefont {Sheehy}\ and\ \citenamefont
  {Radzihovsky}(2007)}]{SHEEHY20071790}%
  \BibitemOpen
  \bibfield  {author} {\bibinfo {author} {\bibfnamefont {D.~E.}\ \bibnamefont
  {Sheehy}}\ and\ \bibinfo {author} {\bibfnamefont {L.}~\bibnamefont
  {Radzihovsky}},\ }\bibfield  {title} {\bibinfo {title} {Bec–bcs crossover,
  phase transitions and phase separation in polarized resonantly-paired
  superfluids},\ }\href
  {https://doi.org/https://doi.org/10.1016/j.aop.2006.09.009} {\bibfield
  {journal} {\bibinfo  {journal} {Annals of Physics}\ }\textbf {\bibinfo
  {volume} {322}},\ \bibinfo {pages} {1790} (\bibinfo {year}
  {2007})}\BibitemShut {NoStop}%
\bibitem [{not()}]{note1}%
  \BibitemOpen
  \bibfield  {title} {\bibinfo {title} {For example, when the
  dispersion is modified to $\xi_{\bm{k}}\rightarrow
  \xi_{\bm{k}}+\delta_1k_{x}^{2}k_{y}^{2}+\delta_2k_{x}k_{y}$, non-zero
  $(\delta_1,\delta_2)$ represent the reduction of the system's rotation
  symmetry from $C_{\infty}$ to $C_{4}$ symmetry with $(\delta_1\neq
  0,\delta_2=0)$, and then further reduced to $C_{2}$ with $(\delta_1\neq
  0,\delta_2\neq 0)$. Under $C_{4}$ symmetry, the theory suggested that the
  system is more likely to form the LO state at low temperatures and for larger
  $\delta_1$ values. Further breaking of the symmetry to enhances the stability
  and stiffness of the LO state, as the LO state itself possesses $C_{2}$
  symmetry}}\href@noop {} {\ }\BibitemShut {NoStop}%
\bibitem [{\citenamefont {Liu}\ \emph {et~al.}(2023)\citenamefont {Liu},
  \citenamefont {Wei}, \citenamefont {He}, \citenamefont {Zhang}, \citenamefont
  {Wang},\ and\ \citenamefont {Wang}}]{liu2023pair}%
  \BibitemOpen
  \bibfield  {author} {\bibinfo {author} {\bibfnamefont {Y.}~\bibnamefont
  {Liu}}, \bibinfo {author} {\bibfnamefont {T.}~\bibnamefont {Wei}}, \bibinfo
  {author} {\bibfnamefont {G.}~\bibnamefont {He}}, \bibinfo {author}
  {\bibfnamefont {Y.}~\bibnamefont {Zhang}}, \bibinfo {author} {\bibfnamefont
  {Z.}~\bibnamefont {Wang}},\ and\ \bibinfo {author} {\bibfnamefont
  {J.}~\bibnamefont {Wang}},\ }\bibfield  {title} {\bibinfo {title} {Pair
  density wave state in a monolayer high-t c iron-based superconductor},\
  }\href@noop {} {\bibfield  {journal} {\bibinfo  {journal} {Nature}\ }\textbf
  {\bibinfo {volume} {618}},\ \bibinfo {pages} {934} (\bibinfo {year}
  {2023})}\BibitemShut {NoStop}%
\bibitem [{\citenamefont {Zhao}\ \emph {et~al.}(2023)\citenamefont {Zhao},
  \citenamefont {Blackwell}, \citenamefont {Thinel}, \citenamefont {Handa},
  \citenamefont {Ishida}, \citenamefont {Zhu}, \citenamefont {Iyo},
  \citenamefont {Eisaki}, \citenamefont {Pasupathy},\ and\ \citenamefont
  {Fujita}}]{zhao2023smectic}%
  \BibitemOpen
  \bibfield  {author} {\bibinfo {author} {\bibfnamefont {H.}~\bibnamefont
  {Zhao}}, \bibinfo {author} {\bibfnamefont {R.}~\bibnamefont {Blackwell}},
  \bibinfo {author} {\bibfnamefont {M.}~\bibnamefont {Thinel}}, \bibinfo
  {author} {\bibfnamefont {T.}~\bibnamefont {Handa}}, \bibinfo {author}
  {\bibfnamefont {S.}~\bibnamefont {Ishida}}, \bibinfo {author} {\bibfnamefont
  {X.}~\bibnamefont {Zhu}}, \bibinfo {author} {\bibfnamefont {A.}~\bibnamefont
  {Iyo}}, \bibinfo {author} {\bibfnamefont {H.}~\bibnamefont {Eisaki}},
  \bibinfo {author} {\bibfnamefont {A.~N.}\ \bibnamefont {Pasupathy}},\ and\
  \bibinfo {author} {\bibfnamefont {K.}~\bibnamefont {Fujita}},\ }\bibfield
  {title} {\bibinfo {title} {Smectic pair-density-wave order in eurbfe4as4},\
  }\href@noop {} {\bibfield  {journal} {\bibinfo  {journal} {Nature}\ }\textbf
  {\bibinfo {volume} {618}},\ \bibinfo {pages} {940} (\bibinfo {year}
  {2023})}\BibitemShut {NoStop}%
\bibitem [{\citenamefont {Wei}\ \emph {et~al.}(2024)\citenamefont {Wei},
  \citenamefont {Liu}, \citenamefont {Ren}, \citenamefont {Wang},\ and\
  \citenamefont {Wang}}]{wei2024observation}%
  \BibitemOpen
  \bibfield  {author} {\bibinfo {author} {\bibfnamefont {T.}~\bibnamefont
  {Wei}}, \bibinfo {author} {\bibfnamefont {Y.}~\bibnamefont {Liu}}, \bibinfo
  {author} {\bibfnamefont {W.}~\bibnamefont {Ren}}, \bibinfo {author}
  {\bibfnamefont {Z.}~\bibnamefont {Wang}},\ and\ \bibinfo {author}
  {\bibfnamefont {J.}~\bibnamefont {Wang}},\ }\bibfield  {title} {\bibinfo
  {title} {Observation of intra-unit-cell superconductivity modulation},\
  }\href@noop {} {\bibfield  {journal} {\bibinfo  {journal} {arXiv preprint
  arXiv:2404.16683}\ } (\bibinfo {year} {2024})}\BibitemShut {NoStop}%
\bibitem [{\citenamefont {Kong}\ \emph {et~al.}(2024)\citenamefont {Kong},
  \citenamefont {Papaj}, \citenamefont {Kim}, \citenamefont {Zhang},
  \citenamefont {Baum}, \citenamefont {Li}, \citenamefont {Watanabe},
  \citenamefont {Taniguchi}, \citenamefont {Gu}, \citenamefont {Lee} \emph
  {et~al.}}]{kong2024observation}%
  \BibitemOpen
  \bibfield  {author} {\bibinfo {author} {\bibfnamefont {L.}~\bibnamefont
  {Kong}}, \bibinfo {author} {\bibfnamefont {M.}~\bibnamefont {Papaj}},
  \bibinfo {author} {\bibfnamefont {H.}~\bibnamefont {Kim}}, \bibinfo {author}
  {\bibfnamefont {Y.}~\bibnamefont {Zhang}}, \bibinfo {author} {\bibfnamefont
  {E.}~\bibnamefont {Baum}}, \bibinfo {author} {\bibfnamefont {H.}~\bibnamefont
  {Li}}, \bibinfo {author} {\bibfnamefont {K.}~\bibnamefont {Watanabe}},
  \bibinfo {author} {\bibfnamefont {T.}~\bibnamefont {Taniguchi}}, \bibinfo
  {author} {\bibfnamefont {G.}~\bibnamefont {Gu}}, \bibinfo {author}
  {\bibfnamefont {P.~A.}\ \bibnamefont {Lee}}, \emph {et~al.},\ }\bibfield
  {title} {\bibinfo {title} {Observation of cooper-pair density modulation
  state},\ }\href@noop {} {\bibfield  {journal} {\bibinfo  {journal} {arXiv
  preprint arXiv:2404.10046}\ } (\bibinfo {year} {2024})}\BibitemShut {NoStop}%
\bibitem [{\citenamefont {Zhang}\ \emph
  {et~al.}(2024{\natexlab{b}})\citenamefont {Zhang}, \citenamefont {Yang},
  \citenamefont {Liu}, \citenamefont {Zhang},\ and\ \citenamefont
  {Fu}}]{zhang2024visualizing}%
  \BibitemOpen
  \bibfield  {author} {\bibinfo {author} {\bibfnamefont {Y.}~\bibnamefont
  {Zhang}}, \bibinfo {author} {\bibfnamefont {L.}~\bibnamefont {Yang}},
  \bibinfo {author} {\bibfnamefont {C.}~\bibnamefont {Liu}}, \bibinfo {author}
  {\bibfnamefont {W.}~\bibnamefont {Zhang}},\ and\ \bibinfo {author}
  {\bibfnamefont {Y.-S.}\ \bibnamefont {Fu}},\ }\bibfield  {title} {\bibinfo
  {title} {Visualizing uniform lattice-scale pair density wave in single-layer
  fese/srtio3 films},\ }\href@noop {} {\bibfield  {journal} {\bibinfo
  {journal} {arXiv preprint arXiv:2406.05693}\ } (\bibinfo {year}
  {2024}{\natexlab{b}})}\BibitemShut {NoStop}%
\bibitem [{\citenamefont {Gu}\ \emph {et~al.}(2023)\citenamefont {Gu},
  \citenamefont {Carroll}, \citenamefont {Wang}, \citenamefont {Ran},
  \citenamefont {Broyles}, \citenamefont {Siddiquee}, \citenamefont {Butch},
  \citenamefont {Saha}, \citenamefont {Paglione}, \citenamefont {Davis} \emph
  {et~al.}}]{gu2023detection}%
  \BibitemOpen
  \bibfield  {author} {\bibinfo {author} {\bibfnamefont {Q.}~\bibnamefont
  {Gu}}, \bibinfo {author} {\bibfnamefont {J.~P.}\ \bibnamefont {Carroll}},
  \bibinfo {author} {\bibfnamefont {S.}~\bibnamefont {Wang}}, \bibinfo {author}
  {\bibfnamefont {S.}~\bibnamefont {Ran}}, \bibinfo {author} {\bibfnamefont
  {C.}~\bibnamefont {Broyles}}, \bibinfo {author} {\bibfnamefont
  {H.}~\bibnamefont {Siddiquee}}, \bibinfo {author} {\bibfnamefont {N.~P.}\
  \bibnamefont {Butch}}, \bibinfo {author} {\bibfnamefont {S.~R.}\ \bibnamefont
  {Saha}}, \bibinfo {author} {\bibfnamefont {J.}~\bibnamefont {Paglione}},
  \bibinfo {author} {\bibfnamefont {J.~S.}\ \bibnamefont {Davis}}, \emph
  {et~al.},\ }\bibfield  {title} {\bibinfo {title} {Detection of a pair density
  wave state in ute2},\ }\href@noop {} {\bibfield  {journal} {\bibinfo
  {journal} {Nature}\ }\textbf {\bibinfo {volume} {618}},\ \bibinfo {pages}
  {921} (\bibinfo {year} {2023})}\BibitemShut {NoStop}%
\bibitem [{\citenamefont {Lee}\ and\ \citenamefont
  {Payne}(1972)}]{PhysRevB.5.923}%
  \BibitemOpen
  \bibfield  {author} {\bibinfo {author} {\bibfnamefont {P.~A.}\ \bibnamefont
  {Lee}}\ and\ \bibinfo {author} {\bibfnamefont {M.~G.}\ \bibnamefont
  {Payne}},\ }\bibfield  {title} {\bibinfo {title} {Pair propagator approach to
  fluctuation-induced diamagnetism in superconductors-effects of impurities},\
  }\href {https://doi.org/10.1103/PhysRevB.5.923} {\bibfield  {journal}
  {\bibinfo  {journal} {Phys. Rev. B}\ }\textbf {\bibinfo {volume} {5}},\
  \bibinfo {pages} {923} (\bibinfo {year} {1972})}\BibitemShut {NoStop}%
\bibitem [{\citenamefont {Hart}\ \emph {et~al.}(2017)\citenamefont {Hart},
  \citenamefont {Ren}, \citenamefont {Kosowsky}, \citenamefont {Ben-Shach},
  \citenamefont {Leubner}, \citenamefont {Brüne}, \citenamefont {Buhmann},
  \citenamefont {Molenkamp}, \citenamefont {Halperin},\ and\ \citenamefont
  {Yacoby}}]{ControlledfinitemomentumpairingandspatiallyvaryingorderparameterinproximitizedHgTequantumwells}%
  \BibitemOpen
  \bibfield  {author} {\bibinfo {author} {\bibfnamefont {S.}~\bibnamefont
  {Hart}}, \bibinfo {author} {\bibfnamefont {H.}~\bibnamefont {Ren}}, \bibinfo
  {author} {\bibfnamefont {M.}~\bibnamefont {Kosowsky}}, \bibinfo {author}
  {\bibfnamefont {G.}~\bibnamefont {Ben-Shach}}, \bibinfo {author}
  {\bibfnamefont {P.}~\bibnamefont {Leubner}}, \bibinfo {author} {\bibfnamefont
  {C.}~\bibnamefont {Brüne}}, \bibinfo {author} {\bibfnamefont
  {H.}~\bibnamefont {Buhmann}}, \bibinfo {author} {\bibfnamefont {L.~W.}\
  \bibnamefont {Molenkamp}}, \bibinfo {author} {\bibfnamefont {B.~I.}\
  \bibnamefont {Halperin}},\ and\ \bibinfo {author} {\bibfnamefont
  {A.}~\bibnamefont {Yacoby}},\ }\bibfield  {title} {\bibinfo {title}
  {Controlled finite momentum pairing and spatially varying order parameter in
  proximitized hgte quantum wells},\ }\href {https://doi.org/10.1038/nphys3877}
  {\bibfield  {journal} {\bibinfo  {journal} {Nature Physics}\ }\textbf
  {\bibinfo {volume} {13}},\ \bibinfo {pages} {87} (\bibinfo {year}
  {2017})}\BibitemShut {NoStop}%
\bibitem [{\citenamefont {Chen}\ \emph {et~al.}(2018)\citenamefont {Chen},
  \citenamefont {Park}, \citenamefont {Gill}, \citenamefont {Xiao},
  \citenamefont {Reig-i Plessis}, \citenamefont {MacDougall}, \citenamefont
  {Gilbert},\ and\ \citenamefont
  {Mason}}]{FinitemomentumCooperpairinginthree-dimensionaltopologicalinsulatorJosephsonjunctions}%
  \BibitemOpen
  \bibfield  {author} {\bibinfo {author} {\bibfnamefont {A.~Q.}\ \bibnamefont
  {Chen}}, \bibinfo {author} {\bibfnamefont {M.~J.}\ \bibnamefont {Park}},
  \bibinfo {author} {\bibfnamefont {S.~T.}\ \bibnamefont {Gill}}, \bibinfo
  {author} {\bibfnamefont {Y.}~\bibnamefont {Xiao}}, \bibinfo {author}
  {\bibfnamefont {D.}~\bibnamefont {Reig-i Plessis}}, \bibinfo {author}
  {\bibfnamefont {G.~J.}\ \bibnamefont {MacDougall}}, \bibinfo {author}
  {\bibfnamefont {M.~J.}\ \bibnamefont {Gilbert}},\ and\ \bibinfo {author}
  {\bibfnamefont {N.}~\bibnamefont {Mason}},\ }\bibfield  {title} {\bibinfo
  {title} {Finite momentum cooper pairing in three-dimensional topological
  insulator josephson junctions},\ }\bibfield  {journal} {\bibinfo  {journal}
  {Nature Communications}\ }\textbf {\bibinfo {volume} {9}},\ \href
  {https://doi.org/10.1038/s41467-018-05993-w} {10.1038/s41467-018-05993-w}
  (\bibinfo {year} {2018})\BibitemShut {NoStop}%
\bibitem [{\citenamefont {Hamidian}\ \emph {et~al.}(2016)\citenamefont
  {Hamidian}, \citenamefont {Edkins}, \citenamefont {Joo}, \citenamefont
  {Kostin}, \citenamefont {Eisaki}, \citenamefont {Uchida}, \citenamefont
  {Lawler}, \citenamefont {Kim}, \citenamefont {Mackenzie}, \citenamefont
  {Fujita} \emph {et~al.}}]{hamidian2016detection}%
  \BibitemOpen
  \bibfield  {author} {\bibinfo {author} {\bibfnamefont {M.}~\bibnamefont
  {Hamidian}}, \bibinfo {author} {\bibfnamefont {S.~D.}\ \bibnamefont
  {Edkins}}, \bibinfo {author} {\bibfnamefont {S.~H.}\ \bibnamefont {Joo}},
  \bibinfo {author} {\bibfnamefont {A.}~\bibnamefont {Kostin}}, \bibinfo
  {author} {\bibfnamefont {H.}~\bibnamefont {Eisaki}}, \bibinfo {author}
  {\bibfnamefont {S.}~\bibnamefont {Uchida}}, \bibinfo {author} {\bibfnamefont
  {M.}~\bibnamefont {Lawler}}, \bibinfo {author} {\bibfnamefont {E.-A.}\
  \bibnamefont {Kim}}, \bibinfo {author} {\bibfnamefont {A.~P.}\ \bibnamefont
  {Mackenzie}}, \bibinfo {author} {\bibfnamefont {K.}~\bibnamefont {Fujita}},
  \emph {et~al.},\ }\bibfield  {title} {\bibinfo {title} {Detection of a
  cooper-pair density wave in bi2sr2cacu2o8+ x},\ }\href@noop {} {\bibfield
  {journal} {\bibinfo  {journal} {Nature}\ }\textbf {\bibinfo {volume} {532}},\
  \bibinfo {pages} {343} (\bibinfo {year} {2016})}\BibitemShut {NoStop}%
\bibitem [{\citenamefont {Cho}\ \emph {et~al.}(2019)\citenamefont {Cho},
  \citenamefont {Bastiaans}, \citenamefont {Chatzopoulos}, \citenamefont {Gu},\
  and\ \citenamefont {Allan}}]{cho2019strongly}%
  \BibitemOpen
  \bibfield  {author} {\bibinfo {author} {\bibfnamefont {D.}~\bibnamefont
  {Cho}}, \bibinfo {author} {\bibfnamefont {K.}~\bibnamefont {Bastiaans}},
  \bibinfo {author} {\bibfnamefont {D.}~\bibnamefont {Chatzopoulos}}, \bibinfo
  {author} {\bibfnamefont {G.}~\bibnamefont {Gu}},\ and\ \bibinfo {author}
  {\bibfnamefont {M.}~\bibnamefont {Allan}},\ }\bibfield  {title} {\bibinfo
  {title} {A strongly inhomogeneous superfluid in an iron-based
  superconductor},\ }\href@noop {} {\bibfield  {journal} {\bibinfo  {journal}
  {Nature}\ }\textbf {\bibinfo {volume} {571}},\ \bibinfo {pages} {541}
  (\bibinfo {year} {2019})}\BibitemShut {NoStop}%
\bibitem [{\citenamefont {Anchenko}\ and\ \citenamefont
  {Zil’Berman}(1969)}]{anchenko1969josephson}%
  \BibitemOpen
  \bibfield  {author} {\bibinfo {author} {\bibfnamefont {Y.}~\bibnamefont
  {Anchenko}}\ and\ \bibinfo {author} {\bibfnamefont {L.}~\bibnamefont
  {Zil’Berman}},\ }\bibfield  {title} {\bibinfo {title} {The josephson effect
  in small tunnel contacts},\ }\href@noop {} {\bibfield  {journal} {\bibinfo
  {journal} {Soviet Phys. JETP}\ }\textbf {\bibinfo {volume} {28}} (\bibinfo
  {year} {1969})}\BibitemShut {NoStop}%
\bibitem [{\citenamefont {Ingold}\ \emph {et~al.}(1994)\citenamefont {Ingold},
  \citenamefont {Grabert},\ and\ \citenamefont {Eberhardt}}]{ingold1994cooper}%
  \BibitemOpen
  \bibfield  {author} {\bibinfo {author} {\bibfnamefont {G.-L.}\ \bibnamefont
  {Ingold}}, \bibinfo {author} {\bibfnamefont {H.}~\bibnamefont {Grabert}},\
  and\ \bibinfo {author} {\bibfnamefont {U.}~\bibnamefont {Eberhardt}},\
  }\bibfield  {title} {\bibinfo {title} {Cooper-pair current through ultrasmall
  josephson junctions},\ }\href@noop {} {\bibfield  {journal} {\bibinfo
  {journal} {Physical Review B}\ }\textbf {\bibinfo {volume} {50}},\ \bibinfo
  {pages} {395} (\bibinfo {year} {1994})}\BibitemShut {NoStop}%
\bibitem [{\citenamefont
  {Takada}(1970{\natexlab{b}})}]{takada1970superconductivity}%
  \BibitemOpen
  \bibfield  {author} {\bibinfo {author} {\bibfnamefont {S.}~\bibnamefont
  {Takada}},\ }\bibfield  {title} {\bibinfo {title} {Superconductivity in a
  molecular field. ii: stability of fulde-ferrel phase},\ }\href@noop {}
  {\bibfield  {journal} {\bibinfo  {journal} {Progress of Theoretical Physics}\
  }\textbf {\bibinfo {volume} {43}},\ \bibinfo {pages} {27} (\bibinfo {year}
  {1970}{\natexlab{b}})}\BibitemShut {NoStop}%
\end{thebibliography}
\end{document}